\documentclass[a4paper,fleqn]{cas-sc}
\usepackage[authoryear]{natbib}

\usepackage{graphicx}
\usepackage{amsmath}
\usepackage{siunitx}
\usepackage{diagbox}
\usepackage{multirow}
\usepackage{upgreek}
\usepackage{rotating}

\usepackage{hyperref}
\hypersetup{colorlinks=true, citecolor=blue, linkcolor=blue}
\usepackage{color}

\def \vR {\mathbf{R}}           

\def \degree {^{\circ}}
\def \inc {I}

\def \ii {{\rm i}}

\def \upomega {\omega}

\def \Ru {R_0}
\def \Grav {\mathcal{G}}
\def \dt {\tau}

\def \klU {k_{2,0}}
\def \KtU {{\cal K}_{k0}}

\def \klk {k_{2,k}}
\def \Ktk {{\cal K}_{0k}}

\def \damp {\tau}

\def \angrot 	{\omega}

\def \obq   {\epsilon}

\begin{document}

\shorttitle{Dynamical evolution of the Uranian satellite system I}    

\shortauthors{S.R.A. Gomes \& A.C.M. Correia}  

\title [mode = title]{Dynamical evolution of the Uranian satellite system I.}  
\title [mode=sub]{From the 5/3 Ariel--Umbriel mean motion resonance to the present}


\author[lab1]{S\'ergio R. A. Gomes}

\cormark[1]


\ead{sergio.ra.gomes@outlook.com}



\author[lab1,lab2]{Alexandre C. M. Correia}


\ead{acor@uc.pt}



\address[lab1]{CFisUC, Departamento de F\'isica, Universidade de Coimbra, 3004-516 Coimbra, Portugal}
\address[lab2]{IMCCE, UMR8028 CNRS, Observatoire de Paris, PSL Université, 77 Av. Denfert-Rochereau, 75014 Paris, France}




\begin{abstract}
Mutual gravitational interactions between the five major Uranian satellites raise small quasi-periodic fluctuations on their orbital elements. 
At the same time, tidal interactions between the satellites and the planet induce a slow outward drift of the orbits, while damping the eccentricities and the inclinations.
In this paper, we revisit the current and near past evolution of this system using a $N-$body integrator, including spin evolution and tidal dissipation with the weak friction model.
We update the secular eigenmodes of the system and show that it is unlikely that any of the main satellites were recently captured into a high obliquity Cassini state.
We rather expect that the Uranian satellites are in a low obliquity Cassini state and compute their values.
We also estimate the current variations in the eccentricities and inclinations, 
and show that they are not fully damped.
We constrain the modified quality factor of Uranus to be $Q_U' = ( 1.2 \pm 0.4 ) \times 10^5$, and that of Ariel to be $ Q_A' = (7 \pm 3) \times 10^4 $.
We find that the system most likely encountered the 5/3 mean motion resonance between Ariel and Umbriel in the past, at about $(0.7 \pm 0.2)$~Gyr ago.
We additionally determine the eccentricities and inclinations of all satellites just after the resonance passage that comply with the current system.
We finally show that, from the crossing of the 5/3 MMR to the present, the evolution of the system is mostly peaceful and dominated by tides raised on Uranus by the satellites.
\end{abstract}


\maketitle


\section{Introduction}

The orbits of the five largest moons of Uranus pose many queries. 
Their proximity with the host planet and the short orbital periods make them a very compact system (Table~\ref{tab:orbital_physical_properties_satellites}), similar to many exoplanetary systems such as Trappist-1 \citep{Gillon_etal_2017}, Kepler-11 \citep{Lissauer_etal_2011}, TOI-178 \citep{Leleu_etal_2021}, or TOI-1136 \citep{Dai_etal_2023}. 
This places the Uranian system as a perfect laboratory to study in detail the formation and evolution of many close-in compact systems.

Uranus spin-axis has an extreme $98\degree$ inclination relatively to its orbital plane.
Such poses the satellite's orbital plane almost perpendicular to the planet's orbital plane. 
Even so, \cite{Laskar_Robutel_1993} showed that the tilt of Uranus is stable and should be considered primordial. 
Moreover, the regular satellites are immune to the solar perturbation \citep[e.g.][see also Sect.~\ref{sec:free_forced_orbital_elements}]{Boue_Laskar_2010} and to the migration effects of the giant planets in the early Solar System \citep{Deienno_etal_2011}. 
Therefore, the long-term evolution of the satellite's orbits can be conducted as an isolated system.

\begin{table*}
    \centering
    \caption{Physical and mean orbital parameters of the five largest Uranian satellites. The masses and orbital parameters are from \citet{Jacobson_2014}, the radius is from \citet{Thomas_1988}, and the second order gravity field, $J_2$, and the tidal Love numbers, $k_2$, are from Table~2 in \citet{Chen_etal_2014}. The fluid Love numbers, $k_{\rm f}$, and the inner structure coefficients, $\zeta$, are obtained from Eqs.\,(\ref{eq:fluid_love_number}) and (\ref{eq:inner_structure_coefficient}), respectively. }
    \label{tab:orbital_physical_properties_satellites}
\renewcommand{\arraystretch}{1.1}
    \begin{tabular}{c|c|c|c|c|c|c}
		\hline       
       & Uranus & Miranda & Ariel & Umbriel & Titania & Oberon\\
       \hline
       Mass ($\rm \times 10^{-10} \, M_{\odot}$)    & $\num{4.365628e5}$ & $0.323997$ & $6.291561$ & $6.412118$ & $17.096471$ & $15.468953$\\
       Radius (km)                                 &  $25\,559$  &$235.8$ & $578.9$ & $584.7$ & $788.4$ & $761.2$ \\
       $J_2$                                       &  $\num{3.5107e-3}$  & $\num{6.10e-3}$ & $\num{1.39e-3}$ &$\num{6.13e-4}$&$\num{1.13e-4}$&$\num{1.48e-5}$\\
       $k_2$                                       &  $\num{0.104}$  & $\num{8.84e-4}$ & $\num{1.02e-2}$ &$\num{7.35e-3}$&$\num{1.99e-2}$&$\num{1.68e-2}$\\
       $k_{\rm f}$                                       &  $\num{0.356}$  & $\num{0.907}$ & $\num{0.862}$ &$\num{1.016}$&$\num{0.899}$&$\num{0.790}$\\
       $\zeta$                                    & \num{0.225}  &  $\num{0.327}$  & $\num{0.320}$ & $\num{0.342}$ &$\num{0.326}$&$\num{0.310}$\\
       Period (day)                                &  $0.718328$  & $1.413480$ & $2.520381$ & $4.144176$ & $8.705883$ & $13.463254$ \\
       $a \, (\Ru)$ & &$5.080715$ & $7.470167$ & $10.406589$ & $17.069604$ & $22.827536$ \\
       $e \, (\times 10^{-3})$ & & $1.35$ & $1.22$ & $3.94$ & $1.23$ & $1.40$ \\
       $\inc \, (\degree)$ & & $4.4072$ & $0.0167$ & $0.0796$ & $0.1129$ & $0.1478$\\
       \hline
    \end{tabular}
    \renewcommand{\arraystretch}{1.0}
    \end{table*}

Natural satellites are believed to be formed in a protoplanetary disk around the host planet \citep[e.g.][]{Peale_1999}, or to be ``free bodies'' that were captured by the host planet \citep[e.g.][]{Singer_1968,Agnor_2006,Jewitt_Haghighipour_2007}. 
Despite the origin of the Uranian satellite system is not yet completely understood \citep[see][for a detailed revision]{Rogoszinski_Hamilton_2020}, the main satellites were likely formed in a circumplanetary disk \citep[e.g.][]{Pollack_etal_1991, Szulagyi_etal_2018, Ishizawa_etal_2019, Inderbitzi_etal_2020, Ida_etal_2020, Rufu_Canup_2022}.
Indeed, data from Voyager 2 shows that the distribution of mass between the satellites and Uranus and their chemical composition are consistent with these bodies having been condensed from a concentric disk, withstanding the idea that the regular moons of Uranus formed in an accretion disk around the planet \citep{Prentice_1986,Pollack_etal_1991}. 
As a consequence, the initial eccentricities and inclinations of the main satellites should have been extremely small.
Yet, the current inclination of Miranda of about $4.3\degree$ is considerably high when compared with  the $ \lesssim 0.1\degree$ inclinations of the remaining satellites \citep{Tittemore_Wisdom_1989,Tittemore_Wisdom_1990,Malhotra_Dermott_1990,Verheylewegen_etal_2013,Cuk_etal_2020}. 

The currently observed eccentricities of the Uranian satellites are also intriguing. 
Although small, they are still abnormally high when tidal dissipation is taken into account \citep[e.g.][]{Squyres_etal_1985}.
In particular, tides are expected to damp the eccentricity of Ariel on a $10^7-10^8$~yr timescale, and the oscillations owing to mutual perturbations between the satellites are not enough to explain its present value \citep{Dermott_Nicholson_1986, Laskar_1986}.

Since the formation of the Solar System, $4.5$ Gyr ago, tidal friction is thought to induce a slow outward differential migration of the main five Uranian satellites \citep[e.g.][]{Peale_1988}. 
The changes on the relative distances between the satellites most likely lead to the crossing of several commensurabilities in the past, where the most recent is believed to be a $5/3$ mean motion resonance (MMR) between Ariel and Umbriel.
Previous studies focussed on the passage through the several low order MMRs possibly encountered by the major Uranian satellites during its orbital evolution \citep{Tittemore_Wisdom_1988, Dermott_etal_1988, Henrard_Sato_1989, Cuk_etal_2020, Gomes_Correia_2023}.
They have shown that the passage through the $5/3$ Ariel-Umbriel MMR, with or without long-term capture, most likely excited the eccentricities and inclinations of the three largest innermost satellites, shaping the current architecture of the system.

\citet{Cuk_etal_2020} observed that if Ariel and Umbriel experienced a long-term capture within the 5/3~MMR, all five moons had their inclinations excited.
The effect on the inclination is particularly significant for Miranda, which could explain its current high value of about $4.3\degree$.
However, the inclination values also acquired by Ariel and Umbriel around $1\degree$ seem difficult to conciliate with the present observed system (Table \ref{tab:orbital_physical_properties_satellites}).
To address this discrepancy, the authors propose a spin-orbit resonance between the precession frequency of Oberon's spin axis and the precession frequency of Umbriel's node as a mechanism to decrease Umbriel's inclination.

An alternative scenario attributes Miranda's high inclination to the passage through the $3/1$~MMR between Miranda and Umbriel \citep{Tittemore_Wisdom_1989,Tittemore_Wisdom_1990}, thus making its crossing somewhere in the past mandatory.
A more plausible scenario is then that Miranda's inclination is excited during the Miranda and Umbriel 3/1~MMR encounter, and then Ariel and Umbriel skip the 5/3~MMR without being captured to prevent any damage on the inclinations.

Despite some studies on the 5/3~MMR passage have been already conducted \citep{Tittemore_Wisdom_1988, Cuk_etal_2020, Gomes_Correia_2023}, the orbital evolution of the main satellites just after this commensurability remain uncertain. 
Far from a resonance, tidal interactions between the satellites and Uranus are expected to dominate the evolution of the system. 
Thus, since the last MMR, the eccentricities and inclinations of the satellites should tend towards a state of equilibrium between the mutual perturbations within the satellites and tidal friction.

In this paper, we revisit the current and near past tidal evolution of the main five Uranian moons.
We aim to understand the current architecture of the system and how the major satellites of Uranus settled into the present state. 
We start by access the present orbital status of the major Uranian satellites, in Sect.~\ref{sec:the_present_system}. 
Then, by adopting the weak friction tidal model, in Sect.~\ref{sec:orbital_evolution}, we study their impact in shaping the present system and derive constraints on the dissipation rates.
Finally, in Sect.\ref{sec:backwards_evolution} we attempt to reconstruct the tidal orbital evolution of the satellites from just after the passage through 5/3~MMR up to the present days.
We discuss our results in Sect.~\ref{sec:conclusion}.
In a companion paper \citep{Gomes_Correia_2024p2}, hereafter Paper~II, we provide an exhaustive view on how the system evaded capture in the 5/3~MMR, allowing it to evolve into the present configuration.


\section{The present system}\label{sec:the_present_system}
\subsection{Orbital architecture}\label{sec:orbital_arquitecture}

\cite{Jacobson_2014} provides the best precise estimation to date of the Uranian satellites' properties.
The masses, $m$, and the radius, $R$, of the planet and the satellites are given in Table~\ref{tab:orbital_physical_properties_satellites}.
We also list the mean orbital elements of the major Uranian satellites, obtained by fitting precessing ellipses to the integrated orbits over 1900–2100.

\begin{table}
    \centering
    \caption{Position and velocity vectors at August $1^{\rm st}$, 1985, on the Earth equatorial frame of the five main satellites of Uranus \citep{Jacobson_2014}.}
    \label{tab:Jacobson_positions_velocities}
    \renewcommand{\arraystretch}{1.1} 
    \begin{tabular}{c|c|c}
    	\hline
        Satellite & Position (km)  &  Velocity ($\text{km.s}^{-1}$)\\
                  & $(X,Y,Z)$ &  $(\dot{X},\dot{Y},\dot{Z})$\\
        \hline 
        Miranda & $-127430.9607930668$  & $-0.422514450329333$  \\
                & $23792.64617013941$   & $-1.271890082631948$  \\
                & $-3464.554580724168$  & $6.552338419694388$   \\
        \hline
        Ariel   & $-185785.2177189803$  & $-0.384730129923274$  \\
                & $42477.81018746200$   & $-1.393752472818678$  \\
                & $-2109.273462727150$  & $5.325004225204424$   \\
        \hline
        Umbriel & $-176566.9475784755$  & $-3.350588413391897$  \\
                & $89016.12833946147$   & $-0.153568184837806$  \\
                & $-3.350588413391897$  & $3.273855499527411$   \\
        \hline
        Titania & $-221240.1941919138$  & $-3.049048602775958$  \\
                &  $145452.9878060127$  & $0.138409610142017$   \\
                & $-346697.1461249496$  & $1.991437563896877$   \\
        \hline
        Oberon  & $-155108.4287158760$  & $-2.962407899779837$  \\
                & $181606.6634411168$   & $0.385864135361727$   \\
                & $-532879.3651011410$  & $0.993694238058708$	\\
        \hline
    \end{tabular}
    \renewcommand{\arraystretch}{1} 
\end{table}

The instantaneous position and velocity vectors of the satellites' orbits, expressed on the Earth's equatorial frame are given in Table~\ref{tab:Jacobson_positions_velocities}, also taken from \citet{Jacobson_2014}.
Departing from the position vector, $\vR=(X,Y,Z)$, and the velocity vector, $\dot{\vR}=(\dot{X},\dot{Y},\dot{Z})$, we compute the corresponding osculating elliptical elements, that is, the semi-major axis, $a$, the eccentricity, $e$, the inclination, $\inc$, the mean longitude, $\lambda$, the longitude of the pericentre, $\varpi$, and the longitude of the ascending node, $\Omega$, given in the Earth equatorial frame.
However, it is more convenient to express the orbital elements in the Uranian equatorial reference frame. 
We start from the Uranus' pole orientation declination and right ascension angles at August $1^{\rm st}$, 1985, respectively
\begin{equation}
	\begin{split}
    \delta & =  15.172593631449473 \degree \ , \\
    RA  & =  77.31003302401596 \degree \ , 
    \end{split}
\end{equation}
obtained from the Appendix B of \cite{Jacobson_2014}. Then, 
by solving \citep[e.g.][]{Smart_1965}
\begin{equation}
    \cos\epsilon=\sin\delta \quad \text{and} \quad  \cos\psi=-\frac{\cos\delta \sin(RA)}{\sin\epsilon} \, ,
\end{equation}  
we compute the angle between the spin axis and the vector normal to the orbital plane, i.e., the obliquity, $\obq$, and the longitude of the Uranus' equator, $\psi$, in the Earth equatorial frame, respectively
\begin{equation}
	\begin{split}
        \obq = & \num{74.82740636855053} \degree \ , \\
        \psi = & \num{167.31003302401598} \degree \ .
	\end{split}
\end{equation}

Finally, using spherical trigonometry to relate the Earth and Uranus equatorial frames \citep[e.g.][]{Smart_1965}, we obtain the osculating elliptical elements of the five major Uranian satellites expressed in the Uranus equatorial frame:
\begin{align}
    \cos{\inc}&=\cos{\epsilon}\cos\inc^\prime+\sin{\epsilon\sin{\inc^\prime}}\cos{(\Omega^\prime-\psi)} \, ,\\
    \cos(\upomega^\prime-\upomega)&=\frac{\cos\epsilon-\cos\inc^\prime\cos\inc}{\sin\inc^\prime\sin\inc} \, ,\\
    \begin{split}
    \cos\Omega&=\cos\Omega^\prime\cos(\upomega^\prime-\upomega)\\
    &-\sin(\Omega^\prime-\psi)\sin(\upomega^\prime-\upomega)\cos\inc^\prime \, ,
    \end{split}
\end{align}
where the primed quantities are expressed in the Earth equatorial frame, the non-primed quantities are expressed in the Uranian equatorial frame, and $\upomega=\varpi-\Omega$.
The final results are given in Table~\ref{tab:elliptical_elements_Uranus_equatorial_frame}.

\begin{table*}
    \centering
    \caption{Osculating elliptical orbital elements of the five main Uranian satellites on the Uranus' equatorial frame at August $1^{\rm st}$, 1985.}
    \label{tab:elliptical_elements_Uranus_equatorial_frame}
    \renewcommand{\arraystretch}{1.1} 
    \begin{tabular}{c|c|c|c}
    \hline
     Satellite  & $a\, (\times 10^{-3})$ (au) & $e \, (\times 10^{-3})$ & $\inc \, (\degree)$   \\
     \hline
     Miranda    & $\num{0.86782980609613350}$ & $\num{1.5143842679902321}$ & $\num{4.4090226441899283}$  \\
     Ariel      & $\num{1.2759486623236246}$ & $\num{1.8054435068308795}$ & $\num{0.016291682978667430}$ \\
     Umbriel    & $\num{1.7774606905474049}$ & $\num{4.1913140312797963}$ & $\num{0.063682350178256561}$  \\
     Titania    & $\num{2.9159539422553278}$ & $\num{2.6914293314397701}$ & $\num{0.12132835838261483}$  \\
     Oberon     & $\num{3.8994952822486825}$ & $\num{1.0218485481986467}$ & $\num{0.15841722443957440}$  \\
     \hline\hline
                & $\lambda \, (\degree)$ & $\varpi \, (\degree)$ & $\Omega \, (\degree)$ \\
     \hline
     Miranda    & $\num{358.8803085315008161}$ & $\num{316.9378996504242991}$ & $\num{32.632119491654876}$  \\
     Ariel      & $\num{359.2250483102935732}$ & $\num{326.0069123828036481}$ & $\num{262.70435953229685}$ \\
     Umbriel    & $\num{316.2968105608910037}$ & $\num{302.7081856231474717}$ & $\num{247.49469574449512}$  \\
     Titania    & $\num{304.3034067438209149}$ & $\num{213.6530622689360825}$ & $\num{353.26006466854500}$  \\
     Oberon     & $\num{289.1109595742070155}$ & $\num{110.6613685473152771}$ & $\num{51.710122333743321}$  \\
     \hline
    \end{tabular}
    \renewcommand{\arraystretch}{1} 
\end{table*}


\subsection{Numerical code}\label{sec:num_code}

In order to run the simulations on the Uranian satellite's system, we use the $N-$body numerical code \texttt{SPINS} \citep{Correia_2018}, that takes into account satellite-satellite interactions, spin dynamics, rotational flattening, and tidal dissipation following the weak friction model. 
The numerical results in this section can also be obtained using the open-source code \texttt{TIDYMESS} \citep{Boekholt_Correia_2023}, with the option \texttt{tidal\_model = 2}, which corresponds to the weak friction tidal model.
Our numerical code also adopts a similar model for rotational flattening and tides as the symplectic integrator \texttt{SIMPL} \citep{Cuk_etal_2016a}, that was used by \citet{Cuk_etal_2020} to study the Uranus' satellite system.

\subsection{Secular modes}\label{sec:secular_modes}

As the Uranian satellites orbit the central planet, mutual gravitational interactions lead to quasi-periodic oscillations of the eccentricities and inclinations over broad periods of time, known as secular variations.
Through semi-analytical simplified models, it is possible to estimate the frequencies of the eigenmodes associated to the eccentricity and inclination of each satellite, known as the secular modes \citep[e.g.][]{Dermott_Nicholson_1986}. 
These frequencies were estimated by \cite{Laskar_1986,Laskar_Jacobson_1987} using the analytical model GUST (General Uranian satellite theory). 
Later, \cite{Malhotra_etal_1989} revisited the results obtained with GUST, incorporating the effects of near first order mean motion resonances, within the LONGSTOP project \citep{Carpino_etal_1987,Milani_etal_1987}.
Since then, the estimations of the satellite masses were updated, as well as the orbital parameters \citep{Jacobson_2014}, and so a recalculation of these frequencies is required.

Departing from the orbital solution from Table~\ref{tab:elliptical_elements_Uranus_equatorial_frame} and the physical properties from Table~\ref{tab:orbital_physical_properties_satellites}, we integrated the system composed by Uranus, Miranda, Ariel, Umbriel, Titania, and Oberon over 100~kyr, disregarding the tidal perturbations.
Through the software \texttt{TRIP} \citep{TRIP}, we performed a frequency analysis of the orbits \citep{Laskar_1990,Laskar_1993} to determine the fundamental frequencies of the eccentricities ($g_k$) and inclinations ($s_k$) of the five major Uranian satellites. 
The results are shown in Table \ref{tab:orbital_frequencies}, alongside the results obtained by \cite{Laskar_Jacobson_1987} and \cite{Malhotra_etal_1989}. 
The subscripts are given in ascending order of the semi-major axes of the satellites.
We observe that there is a good agreement between all frequency values in Table \ref{tab:orbital_frequencies}.
We thus conclude that the orbital secular modes are neither very sensitive to the improvements of the physical parameters performed by \cite{Jacobson_2014} nor to the inclusion of the spins of all bodies in the system.
For completeness, we also determine the amplitudes of the linear Lagrange-Laplace solution of $z = e \, \mathrm{e}^{\ii  \varpi } $ (Table~\ref{tab:secular_modes_ecc}) and $\eta = \sin (\inc/2) \, \mathrm{e}^{\ii  \Omega} $ (Table~\ref{tab:secular_modes_inc}) for the five main satellites of Uranus, also obtained through frequency analysis.

\begin{table}
    \centering
    \caption{Secular modes of Miranda ($g_1,s_1$), Ariel ($g_2,s_2$), Umbriel ($g_3,s_3$), Titania ($g_4,s_4$), and Oberon ($g_5,s_5$), obtained through frequency analysis of \texttt{GUST} \citep{Laskar_Jacobson_1987}, \texttt{LONGSTOP} \citep{Malhotra_etal_1989}, and the orbital evolution of the \texttt{SPINS} $N-$body code over 100~kyr (Sect.~\ref{sec:secular_modes}).}
    \label{tab:orbital_frequencies}
    \begin{tabular}{c|c|c|c c c|c|c|c}
    	\cline{1-4} \cline{6-9}
         & GUST & LONGSTOP & SPINS   & & & GUST & LONGSTOP & SPINS \\
         & (deg/yr) & (deg/yr) & (deg/yr) & & & (deg/yr) & (deg/yr) & (deg/yr) \\
        \cline{1-4} \cline{6-9}
        $g_1$ & $20.082$ & $20.117$ & $20.0417$ & & $s_1$ & $-20.309$  & $-20.340$   & $-20.2408$	\\
        $g_2$ & $6.217$  & $6.186$  & $6.2271$  & & $s_2$ & $-6.287$   & $-6.239$    & $-6.2363$	\\
        $g_3$ & $2.865$  & $2.848$  & $2.8506$  & & $s_3$ & $-2.836$   & $-2.790$    & $-2.7674$	\\
        $g_4$ & $2.079$  & $2.086$  & $2.1652$  & & $s_4$ & $-1.843$   & $-1.839$    & $-1.8309$	\\
        $g_5$ & $0.386$  & $0.410$  & $0.4063$  & & $s_5$ & $-0.259$   & $-0.269$    & $-0.2677$ 	\\
        \cline{1-4} \cline{6-9}
    \end{tabular}
\end{table}

\begin{table*}
     \centering
     \caption{Secular modes and amplitudes of the linear Lagrange-Laplace solution of $z = e \, \mathrm{e}^{\ii  \varpi} $ for the five main satellites of Uranus, obtained through frequency analysis of the orbital evolution of the numerical code over 100~kyr (Sect.~\ref{sec:secular_modes}).}
     \begin{tabular}{c |r |r |r |r |r |r |r}
     \hline
		& & & Miranda & Ariel & Umbriel & Titania & Oberon \\
freq.		& (deg/yr)	 & $\phi_k$ (deg) & $A_k\,(\times10^{-6})$ & $A_k\,(\times10^{-6})$ & $A_k\,(\times10^{-6})$ & $A_k\,(\times10^{-6})$ & $A_k\,(\times10^{-6})$  \\
     \hline
     $g_1$ 	& \num{20.042}&   \num{313.126} & \num{    1299.535} & \num{       2.966} &  &  &  \\
	 $g_2$ 	& \num{6.227} &	  \num{341.119} & \num{      59.866} & \num{    1131.602} & \num{    211.089} & \num{   56.805} & \num{    17.024} \\
	 $g_3$ 	& \num{2.851} &   \num{314.931} & \num{      59.659} & \num{     812.716} & \num{   3628.894} & \num{     397.368} & \num{     158.321} \\        
	 $g_4$ 	& \num{2.165} &   \num{237.733} & \num{       8.364} & \num{     101.800} & \num{    443.203} & \num{     906.305} & \num{     788.069} \\                
	 $g_5$ 	& \num{0.406} &   \num{193.770} & \num{       6.277} & \num{      59.575} & \num{    238.199} & \num{    1260.953} & \num{    1424.890} \\  
     \hline
     \end{tabular}
     \label{tab:secular_modes_ecc}
 \end{table*}

\begin{table*}
     \centering
     \caption{Secular modes and amplitudes of the linear Lagrange-Laplace solution of $\eta = \sin (\inc/2) \mathrm{e}^{\ii  \Omega} $ for the five main satellites of Uranus, obtained through frequency analysis of the orbital evolution of the numerical code over 100~kyr (Sect.~\ref{sec:secular_modes}).}
     \begin{tabular}{c |r |r |r |r |r |r |r}
     \hline
		& & & Miranda & Ariel & Umbriel & Titania & Oberon \\
 freq.			& (deg/yr) & $\phi_k$ (deg) & $A_k\,(\times10^{-6})$ & $A_k\,(\times10^{-6})$ & $A_k\,(\times10^{-6})$ & $A_k\,(\times10^{-6})$ & $A_k\,(\times10^{-6})$ \\
     \hline
     $s_1$ 	& \num{-20.241}	& \num{212.624} & \num{   38463.617} & \num{     113.341} & \num{      10.661} & \num{       1.385} & \num{       0.424} \\     
     $s_2$ 	& \num{-6.236}  & \num{21.667}  & \num{       7.403} & \num{     113.440} & \num{      23.628} & \num{       0.341} & \num{        0.002} \\
     $s_3$ 	& \num{-2.767}  & \num{214.343} & \num{      13.653} & \num{     145.813} & \num{     587.441} & \num{      70.654} & \num{      16.465} \\
     $s_4$ 	& \num{-1.831}	& \num{284.152} & \num{       8.891} & \num{      82.185} & \num{     306.054} & \num{     655.985} & \num{     541.981} \\
     $s_5$ 	& \num{-0.268}	& \num{29.600}  & \num{       3.026} & \num{      72.212} & \num{     252.155} & \num{     949.090} & \num{    1143.359} \\ 
         \hline
     \end{tabular}
     \label{tab:secular_modes_inc}
 \end{table*}


\subsection{Cassini states}\label{sec:spin_orbit_resonances}

Tidal effects are expected to push the obliquity of the satellites, $\obq$, towards zero \citep[e.g.][]{Hut_1980, Correia_2009}, unless an additional perturbation maintains a non-zero obliquity, known as Cassini state \citep[e.g.][]{Colombo_1966, Ward_1975, Correia_2015}. 
Some of these states correspond to stable equilibria and become the end point of tidal dissipation.
More precisely, by decomposing the orbital solution for the inclination in the secular modes (Table~\ref{tab:secular_modes_inc}),
\begin{equation}\label{eq:inc_sec_modes}
\eta = \sin (\inc/2) \, \mathrm{e}^{\ii  \Omega} = \sum_k A_k \mathrm{e}^{\ii  (s_k t + \phi_k) }  \ ,
\end{equation}
we can write for each satellite \citep[e.g.][]{Laskar_Robutel_1993, Correia_Laskar_2003I, Levrard_etal_2007},
\begin{equation}\label{eq:obliquity}
\dot \obq = \sum_k 2 s_k A_k \cos (\psi - s_k t + \phi_k)  - T_\obq \sin \obq \ ,
\end{equation}
\begin{equation}\label{eq:precession}
\dot \psi = - \alpha \cos \obq + \cot \obq \sum_k 2 s_k A_k \sin (\psi - s_k t + \phi_k) \ ,
\end{equation}
where $\alpha$ is the precession constant, $s_k$ are the secular orbital nodal precession frequencies (Table~\ref{tab:orbital_frequencies}), $A_k$ and $\phi_k$ are the respective amplitude and phase of the orbital forcing (Table~\ref{tab:secular_modes_inc}), and $T_\obq$ is the tidal torque \citep[e.g.][]{Correia_Valente_2022}.
For small amplitude variations of the inclination, we have $|A_k| \ll 1$ (Table~\ref{tab:secular_modes_inc}), and so the stable Cassini equilibria for the spin are given by \citep[e.g.][]{Ward_Hamilton_2004, Levrard_etal_2007}
\begin{equation}\label{eq:spin_precession_Millholland_1}
\tan \obq_1 \approx \frac{2 A_k}{1+\alpha / s_k} \ , 
\end{equation}
and
\begin{equation}\label{eq:spin_precession_Millholland_2}
\cos \obq_2 \approx - \frac{s_k}{\alpha} \ .
\end{equation}

For a given satellite in a near circular orbit, we have \citep[e.g.][]{Goldstein_1950, Correia_Rodriguez_2013}
\begin{equation}\label{eq:precession_constant}
    \alpha \approx \frac{3}{2} \frac{n^2}{ \angrot} \frac{ C - A}{C}
    = \frac{3}{2 } \frac{n^2}{ \angrot} \left( J_{2} + 2 C_{22} \right) \zeta^{-1} \, ,
\end{equation}
where $n$ is the orbital mean motion, $\angrot$ is the rotational angular velocity, $C$ and $A$ are the maximal and minimal moments of inertia, respectively, $J_{2}$ and $C_{22}$ are second order gravity field coefficients, and $\zeta$ is an inner structure coefficient.
For a synchronous satellite ($\angrot=n$), the gravity field coefficients and $\zeta$ can be related to the fluid potential Love number, $k_{\rm f}$, as \citep[e.g.][]{Correia_Rodriguez_2013}
\begin{equation}\label{eq:fluid_love_number}
J_{2} =  k_{\rm f} \frac{5}{6} \left(\frac{m_0}{m} \right) \left( \frac{R}{a} \right)^3 \ , 
\end{equation}
\begin{equation}
C_{22} = \frac{k_{\rm f} }{4} \left(\frac{m_0}{m} \right) \left( \frac{R}{a} \right)^3 = \frac{3}{10} J_{2}  \ ,
\end{equation}
and \citep[e.g.][]{Jeffreys_1976}
\begin{equation}\label{eq:inner_structure_coefficient}
    \zeta = \frac{ C}{m R^2} = \frac{2}{3}\left(1-\frac{2}{5}\sqrt{\frac{4-k_{\rm f}}{1+k_{\rm f}}}\right) \, ,
\end{equation}
where $m_0$ is the mass of Uranus, and $m$ and $R$ are the mass and the radius of the satellite, respectively.
From the $J_2$ values in Table~\ref{tab:orbital_physical_properties_satellites}, estimated by \citet{Chen_etal_2014}, we can obtain $k_{\rm f}$ and then $\zeta$ for all satellites.
Adopting the present semi-major axes of the satellites (Table~\ref{tab:orbital_physical_properties_satellites}), with Eq.\,(\ref{eq:precession_constant}) we finally estimate the precession constant (see Table~\ref{tab:spin_axis_precession_constant}).

Using the same method performed to estimate the secular modes (Sect.~\ref{sec:secular_modes}), the precession rate (Eq.\,(\ref{eq:precession})) can also be obtained using a frequency analysis.
From the results of the integration of the Uranian system obtained in Sect.~\ref{sec:secular_modes}, we used \texttt{TRIP} to analyse the precession frequency of the satellites when the obliquity is nearly zero.
This comparison is important, because the precession rate is also influenced by the perturbations of the remaining satellites in the system (Eq.\,(\ref{eq:precession})).
The numerical results are also shown in Table~\ref{tab:spin_axis_precession_constant}.
We observe that there is a good agreement between the precession constant, $\alpha$ (Eq.\,(\ref{eq:precession_constant})), and the main frequency of the precession angle, $\dot \psi$ (Eq.\,(\ref{eq:precession})), meaning that the precession of the satellites is indeed dominated by the gravitational torque of Uranus (term in $\alpha \cos \obq $).

\begin{table}
    \centering
    \caption{Comparison between the precession constant of a synchronous rotating satellite, $\alpha$ (Eq.\,(\ref{eq:precession_constant})), and the numerical precession rate, $\dot \psi$ (Eq.\,(\ref{eq:precession})), obtained through frequency analysis with the \texttt{SPINS}  $N-$body code (Sect.~\ref{sec:num_code}) for a near zero obliquity.}
    \label{tab:spin_axis_precession_constant}
    \renewcommand{\arraystretch}{1.1} 
    \begin{tabular}{c|c|c}
    	\hline
        Satellite & $\alpha$ (deg/yr) &  $|\dot \psi|$ (deg/yr)\\
        \hline
        Miranda 	& $4164.7182$	& $4078.5164$	\\
        Ariel   	& $542.6365$	& $540.3787$    \\
        Umbriel 	& $136.4137$	& $136.1541$    \\
        Titania 	& $12.5519$     & $12.5696$     \\
        Oberon  	& $3.3546$      & $3.1287$ 		\\
	\hline    
    \end{tabular}
    \renewcommand{\arraystretch}{1} 
\end{table}

During the passage through the 5/3~MMR between Ariel and Umbriel, \citet{Cuk_etal_2020} observed that all five moons had their inclinations excited.
The effect on the inclination is particularly significant in Miranda because it has the smallest mass, which could explain its current high value.
However, after leaving the resonance, the remaining four moons are all left with inclination values higher than the current ones.
\citet{Cuk_etal_2020} then propose that the inclination of Umbriel can be lowered if its node is involved in a secular spin-orbit resonance with the spin of Oberon, that is, if $\alpha_\mathrm{oberon} \approx s_3$, which corresponds to Cassini state~2. 

In Tables~\ref{tab:cassini_state_1} and \ref{tab:spin_orbit_angles}, we compute the equilibrium obliquities for all satellites and secular modes in Cassini state~1 (Eq.\,(\ref{eq:spin_precession_Millholland_1})) and Cassini state~2 (Eq.\,(\ref{eq:spin_precession_Millholland_2})), respectively.
We observe that most resonant states (state~2) can only occur for obliquities higher than $60^\circ$.
These high obliquities are unlikely and difficult to explain for tidally evolved close-in bodies such as the satellites of Uranus \citep[e.g.][]{Correia_Laskar_2003I,Levrard_etal_2007,Fabrycky_etal_2007}.
Moreover, in the scenario evoked by \citet{Cuk_etal_2020}, the initial obliquity must be close to zero and then grow to the high value as the inclination is damped owing to angular momentum exchanges \citep[see also][]{Correia_etal_2016}.
As a result, it is only possible to trigger this mechanism when $\alpha \sim s_k$.

The only case where this scenario seems possible is for Oberon, whose spin axis can resonate with $s_3$ and develop an obliquity of about $28^\circ$ (Table~\ref{tab:spin_orbit_angles}), which is exactly the case reported by \citet{Cuk_etal_2020}.
However, the non-zero obliquity would generate an excess of tidal energy dissipated within Oberon.
Using the weak friction tidal model, we have for the tidal heat flux \citep[e.g][]{Levrard_etal_2007, Correia_Valente_2022}
\begin{equation}
	\frac{\dot{E}}{4\pi R^2}=\frac{3 n}{4 \pi} \frac{k_2}{Q}\frac{\Grav m_0^2}{R^3} \left(\frac{R}{a}\right)^6\frac{7e^2 + \sin^2{\obq}}{1+\cos^2{\obq}} \, .
\end{equation}
Adopting $Q/k_2 = 10^4$, we obtain a heat flux of about $ 10^{-5} \, \rm mW m^{-2}$ for a small obliquity of $0.1^\circ$, while we get a heat flux of about $7 \, \rm mW m^{-2}$ for $\obq=28^\circ$. 
The heat fluxes necessary to create the geological features observed on Titania's surface, which is the most similar moon to Oberon, range between $5-12 \, \rm mW m^{-2}$ \citep{Beddingfield_etal_2023}. 
Similarly, the heat fluxes required to form features on Ariel's surface range between $1-16 \, \rm mW m^{-2}$ \citep{Beddingfield_etal_2022}. 
Therefore, we expect that a non-zero obliquity of Oberon around $28^\circ$ would leave recent geological traces on its surface.
Unfortunately, data from Voyager~2 have shown that, among the five regular moons, Oberon presents one of the most cratered and ancient surfaces \citep{Smith_etal_1986, Plescia_1987, Avramchuk_etal_2007, Kirchoff_etal_2022, Bottke_etal_2024}. 
In addition, we would also need that other satellites simultaneously enter in resonance with $s_2$, $s_4$, and $s_5$, in order to damp the inclinations of Ariel, Titania, and Oberon, respectively.
Therefore, the scenario proposed by \cite{Cuk_etal_2020} does not seem a suitable explanation to decrease the inclinations of the satellites after the passage through the 5/3~MMR between Ariel and Umbriel.
We rather expect that the obliquities of the Uranian satellites present low obliquities, corresponding to Cassini state~1 (Table~\ref{tab:cassini_state_1}).

\begin{table}
    \centering
   \caption{Equilibrium obliquities for Cassini state~1 (Eq.\,(\ref{eq:spin_precession_Millholland_1})) for each individual value of $s_k$ and $A_k$ from Table~\ref{tab:secular_modes_inc}.}
    \begin{tabular}{c|c|c|c|c|c}
    	\hline
        \multirow{2}{1cm}{Satellite} & \multicolumn{5}{c}{Cassini state~1 (deg)} \\ \cline{2-6}      & $s_1$ & $s_2$ & $s_3$ & $s_4$ & $s_5$ \\
        \hline
        Miranda 	& ${2.2\times10^{-2}}$ & $\num{1.3e-06}$ & $\num{1.1e-06}$ & $\num{4.6e-07}$ & $\num{2.3e-08}$ \\
        Ariel   	& ${5.1\times10^{-4}}$ & $\num{1.5e-04}$ & $\num{8.6e-05}$ & $\num{3.2e-05}$ & $\num{4.1e-06}$  \\
        Umbriel 	& $\num{2.1e-04}$ & $\num{1.3e-04}$ & ${1.4\times10^{-3}}$ & $\num{4.8e-04}$ & $\num{5.7e-05}$ \\
        Titania 	& $\num{4.2e-04}$ & $\num{3.9e-05}$ & $\num{2.3e-03}$ & ${1.3\times10^{-2}}$ & $\num{2.4e-03}$ \\
        Oberon  	& $\num{5.7e-05}$ & $\num{2.2e-05}$ & $\num{1.4e-02}$ & ${8.8\times10^{-2}}$ & $\num{1.2e-02}$ \\
	\hline 
    \end{tabular}
    \label{tab:cassini_state_1}
\end{table}

\begin{table}
    \centering
  \caption{Equilibrium obliquities for Cassini state~2 (Eq.\,(\ref{eq:spin_precession_Millholland_2})) for each individual value of $s_k$ from Table~\ref{tab:secular_modes_inc}.}
    \begin{tabular}{c|c|c|c|c|c}
    	\hline
        \multirow{2}{1cm}{Satellite} & \multicolumn{5}{c}{Cassini state~2 (deg)} \\ \cline{2-6}      & $s_1$ & $s_2$ & $s_3$ & $s_4$ & $s_5$ \\
        \hline
        Miranda 	& $89.9$ & $89.9$ & $90.0$ & $90.0$ & $90.0$ \\
        Ariel   	& $87.9$ & $89.3$ & $89.7$ & $89.8$ & $90.0$ \\
        Umbriel 	& $81.5$ & $87.4$ & $88.8$ & $89.2$ & $89.9$ \\
        Titania 	& -		 & $60.2$ & $77.3$ & $81.6$ & $88.8$ \\
        Oberon  	& -		 & -	  & $27.7$ & $54.1$ & $85.1$ \\
	\hline 
    \end{tabular}
    \label{tab:spin_orbit_angles}
\end{table}


\section{Orbital evolution}\label{sec:orbital_evolution}

Tidal effects arise from differential and inelastic deformations of an extended body owing to the gravitational effect of a perturber.
The resulting dissipation of tidal energy modifies the spin and the orbit of the extended body.
The final outcome corresponds to circular orbits, zero obliquity, and synchronous rotation \citep[e.g.][]{Hut_1980}. 
For a planet-satellite system, tides raised by the planet on the satellite are much stronger than tides raised by the satellite on the planet.
Therefore, the satellite is expected to reach synchronous rotation on a short timescale, while the planetary spin continues to evolve slowly \citep[e.g.][]{Correia_2009}.

The dissipation of the mechanical energy of tides inside the extended body can be characterised by a time delay, $\dt$, between the maximal deformation and the initial perturbation.
As in many previous studies, we adopt here the weak friction tidal model \citep[e.g.][]{Singer_1968, Alexander_1973, Mignard_1979}, which assumes a constant and small-time delay (Sect.~\ref{sec:num_code}).
For a better understanding of the orbital evolution of the satellites, we show here the equations of motion for zero obliquity ($\obq=0$), small eccentricity, and small inclination for this tidal model \citep[e.g.][]{Correia_etal_2011, Correia_Valente_2022}:

\begin{equation}\label{eq:semi_major_axis_tidal_evolution_full}
   \frac{\dot{a}_k}{a_k}  \approx \KtU \left(\left(2+27 e_k^2-\inc_k^2\right)\frac{\angrot_0}{n_k}-\left(2+46e_k^2\right)\right)  
    + \Ktk \left(\left(2+27e_k^2-\inc_k^2\right)\frac{\angrot_k}{n_k} -\left(2+46e_k^2\right)\right)  \, ,
\end{equation}
\begin{equation}\label{eq:eccentricitu_tidal_evolution_full}
    \frac{\dot{e}_k}{e_k}  \approx  \KtU \left(\frac{11}{2}\frac{\angrot_0}{n_k}-9\right) 
     + \Ktk \left(\frac{11}{2} \frac{\angrot_k}{n_k}-9\right) \ ,
\end{equation}
\begin{equation}\label{eq:inclination_tidal_evolution_full}
 \frac{\dot{\inc}_k}{\inc_k} \approx - \KtU \frac{\angrot_0}{n_k}  - \Ktk \left(\frac{\angrot_k}{n_k}-1\right) \ ,
\end{equation}
with
\begin{equation}\label{tidal_coeficients}
\KtU = \klU \frac{3 \Grav m _k^2 R_0^5}{\beta_k a_k^8} \dt_0 
\ , \quad \mathrm{and} \quad
\Ktk = \klk \frac{3 \Grav m _0^2 R_k^5}{\beta_k a_k^8} \dt_k \ ,
\end{equation}
where $\Grav$ is the gravitational constant, $\beta_k = m_k m_0/ (m_k + m_0) \approx m_k$ is the reduced mass,
$\klk$ is the tidal second Love number, and $\dt_k$ is the time delay.
The subscripts $_0$ and $_k$ pertain to Uranus and the satellite with index $k$, respectively.

Despite Miranda's relatively high inclination, the satellites present very small eccentricities and inclinations (Table~\ref{tab:elliptical_elements_Uranus_equatorial_frame}) and they are expected to present synchronous rotation, that is, ${\obq_k \approx e_k \approx \inc_k \approx 0}$ and ${\angrot_k / n_k \approx 1}$.
Therefore, the orbital evolution can be further simplified as
\begin{equation}\label{eq:semi_major_axis_tidal_evolution}
   \frac{\dot{a}_k}{a_k} \approx 2 \KtU \left(\frac{\angrot_0}{n_k}-1\right) \ ,
\end{equation}
\begin{equation}\label{eq:eccentricitu_tidal_evolution}
    \frac{\dot{e}_k}{e_k} \approx  \KtU \left(\frac{11}{2} \frac{\angrot_0}{n_k}-9\right) - \frac72 \Ktk \ ,
\end{equation}
\begin{equation}\label{eq:inclination_tidal_evolution}
 \frac{\dot{\inc}_k}{\inc_k} \approx - \KtU \frac{\angrot_0}{n_k} \ .
\end{equation}

\subsection{Outward migration}
\label{outmigr}

For the main satellites of Uranus, we always have $\angrot_0/n_k > 1$ (Table~\ref{tab:orbital_physical_properties_satellites}), i.e., all satellites orbit super-synchronously.
We thus conclude that the semi-major axes increase (Eq.\,(\ref{eq:semi_major_axis_tidal_evolution})) and the orbits expand outwards.
The evolution timescale depends on the time delay $\dt_0$, which is related to the tidal dissipation inside Uranus.

A commonly used dimensionless quantity to measure the tidal dissipation is given by the quality factor \citep[e.g.][]{Correia_Valente_2022}, 
\begin{equation}\label{eqQfactor}
    Q_0 \approx \frac{1}{2\angrot_0\dt_0} \quad \mathrm{and} \quad Q_k \approx \frac{1}{n_k\dt_k} \ .
\end{equation}
\citet{Tittemore_Wisdom_1990} constrained the interval of $Q_0$ to be between $11\,000-39\,000$ by studying the likelihood of resonance crossing in the Uranian system.
On one hand, they noticed that the 2/1~MMR between Ariel and Umbriel could not be crossed because the system is unable to evade it, and, on the other hand, the 3/1~MMR between Miranda and Umbriel should have been crossed such that we can explain the currently observed high inclination of Miranda \citep{Tittemore_Wisdom_1989}.

Using the same approach, we recalculate the acceptable values for $Q_0$ using the weak friction tidal model and the updated parameters from Table~\ref{tab:orbital_physical_properties_satellites}.
We start by estimating the current total angular momentum of the system projected on the direction of the spin axis of Uranus:
\begin{equation}\label{eq:Sigma}
	\Sigma = C_0 \angrot_0 + \sum_{k=1}^5 \left( C_k\angrot_k \cos \obq_k + \beta_k\sqrt{\mu_k a_k (1-e_k^2)} \cos \inc_k \right) \, ,
\end{equation}
with $\mu_k=\Grav(m_0+m_k)$ and $C_k$ is the moment of inertia of $m_k$ (see Sect.~\ref{sec:spin_orbit_resonances}).
Adopting the current $\angrot_0$ and semi-major axes' values along with $\obq_k = e_k=\inc_k=0$, and $\angrot_k/n_k=1$, we obtain
\begin{equation}\label{SigmaTOT}
\Sigma=\num{9.446071E-010} \ \mathrm{M_\odot \, au^2 \, yr^{-1}} \ . 
\end{equation}
We then integrate Eq.\,(\ref{eq:semi_major_axis_tidal_evolution}) backwards for the five largest satellites of Uranus with $\klU = 0.104$ \citep{Gavrilov_Zharkov_1977}.
The evolution of $\angrot_0$ is obtained using the conservation of $\Sigma$ and considering that the satellites remain synchronous (Eq.\,(\ref{eq:Sigma})).
The total integration time is chosen to be 4.5~Gyr, the estimated age of the Solar System.
Indeed, age determination studies based on comparisons of crater populations and cratering rates estimate that the surfaces of Uranus' satellites are older than 4~Gyr \citep[e.g.][]{Kirchoff_etal_2022, Bottke_etal_2024}. 

\begin{figure*}
    \centering
    \includegraphics[width=\textwidth]{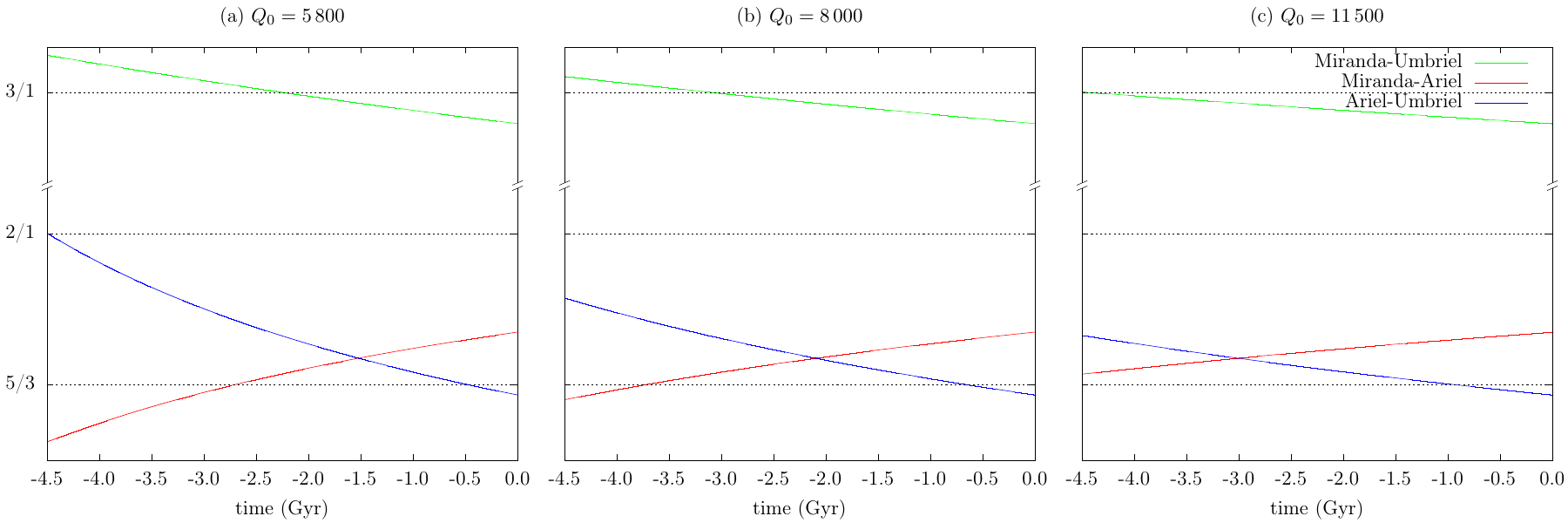}
    \caption{Backwards tidal evolution of the semi-major axes of Miranda, Ariel, and Umbriel using Eq.\,(\ref{eq:semi_major_axis_tidal_evolution}) with $Q_0=5\,800$ (a), $Q_0=8\,000$ (b), and $Q_0=11\,500$ (c). The green line gives the mean motion ratio between Miranda and Umbriel, the red line gives the same ratio for Miranda and Ariel, and the blue line is for Ariel and Umbriel. The dashed line gives the position of the nominal 3/1, 2/1, and 5/3~MMRs.}
    \label{fig:MMR_crossing}
\end{figure*}

In Fig.~\ref{fig:MMR_crossing}, we show the evolution of the mean motion ratios of Miranda, Ariel, and Umbriel.
We observe that for $Q_0<5\,800$, the $2/1$~MMR between Ariel and Umbriel is crossed (see Fig.\,\ref{fig:MMR_crossing}\,a), while for $Q_0>11\,500$, the $3/1$~MMR between Miranda and Umbriel is not crossed (see Fig.\,\ref{fig:MMR_crossing}\,c).
We thus constrain $Q_0$ within $5\,800-11\,500$, or conversely,
\begin{equation}\label{Qestimation}
Q_0 = ( 8.6 \pm 2.9 ) \times 10^3 \ .
\end{equation}
Alternatively, we can compute the modified quality factor, which is independent of the uncertainty in the Love number \citep[e.g.][]{Ogilvie_Lin_2007}
\begin{equation}\label{Qeffestimation}
Q_0' = \frac{3 Q_0}{2 \klU} = ( 1.2 \pm 0.4 ) \times 10^5 \ .
\end{equation}

From the $Q_0$ limits above, we can also estimate that the 5/3~MMR between Ariel and Umbriel was crossed within $460 - 920$~Myr ago, that is,
\begin{equation}\label{time_estimation}
T_{5/3} = - (  0.69 \pm 0.23 ) \, \mathrm{Gyr} \ .
\end{equation}

As in \citet{Gomes_Correia_2023}, we adopt $Q_0=8\,000$ as a suitable value for the tidal dissipation in Uranus for the remainder of this study and also in Paper~II.
From Eq.\,(\ref{eqQfactor}), we compute $\dt_0 \approx 0.62$~s. 
Actually, the exact dissipation rate depends on the product $\klU \, \dt_0$ (Eq.\,(\ref{eq:semi_major_axis_tidal_evolution})), which translates into
\begin{equation}\label{k2sQ}
    \frac{\klU}{Q_0} = \num{1.3e-5} \quad \Leftrightarrow \quad \klU \, \dt_0 = 0.064~\mathrm{s} \ .
\end{equation}
In Fig.~\ref{fig:MMR_crossing}\,b, we also show the backwards tidal evolution of the mean motion ratios for  $Q_0=8\,000$.
From Eq.\,(\ref{eq:semi_major_axis_tidal_evolution}) we then estimate that the 5/3~MMR between Ariel and Umbriel was crossed about 640~Myr ago, with 
\begin{equation}\label{nominal:sma}
    a_2/\Ru=7.39054 \,, \quad a_3/\Ru=10.38909 \,.
\end{equation}

\subsection{Eccentricity damping}\label{sec:ecc_damp}

Contrarily to the semi-major axis, the evolution of the eccentricity (Eq.\,(\ref{eq:eccentricitu_tidal_evolution})) has two different contributions, one from tides raised in Uranus (term in $\KtU$) and another from tides raised in the satellite (term in $\Ktk$).

In order to estimate $\Ktk$, we adopt the tidal Love numbers listed in Table~\ref{tab:orbital_physical_properties_satellites} \citep[for more details see][]{Chen_etal_2014}.
For icy moons, we expect $Q_k \sim 10^2 - 10^3$ (e.g. \cite{Murray_Dermott_1999}).
We thus have $\KtU / \Ktk \ll 1$, and so the eccentricity evolution (Eq.\,(\ref{eq:eccentricitu_tidal_evolution})) is controlled by tides raised on the satellites,
\begin{equation}\label{eq:eccentricitu_tidal_evolution_simple}
    \frac{\dot{e}_k}{e_k} \approx - \frac72 \Ktk \ .
\end{equation}
We conclude that tides damp the eccentricity in a timescale
\begin{equation}
\label{damp:ecc}
    \damp_{e,k} \approx \frac{2}{7 \Ktk} \ .
\end{equation}

Previous studies show that Ariel and Umbriel exit the $5/3$ MMR with average eccentricities in the order of $e_k\sim 10^{-2}$ \citep{Tittemore_Wisdom_1988,Cuk_etal_2020}, a result that is also confirmed by our numerical simulations in Paper~II. 
The current mean eccentricities of Ariel and Umbriel are $e_2 \sim e_3 \sim 10^{-3}$ (Table~\ref{tab:free_eccentricities_inclinations}).
Since the 5/3~MMR is expected to have been crossed about 640~Myr ago (Sect.~\ref{outmigr}), from Eq.\,(\ref{eq:eccentricitu_tidal_evolution_simple}) we estimate that $Q_2 \approx 500$ is the best suited value in order to damp the eccentricity of Ariel by an order of magnitude. Considering the proposed limits for the $Q_0$ value (Eq.\,(\ref{Qestimation})), and following the same line of reasoning, the $Q_2$ value can also be constrained to 
\begin{equation}
Q_2=500 \pm 200 \ .
\end{equation}
Alternatively, using the $k_{2,2}$ value (Table~\ref{tab:orbital_physical_properties_satellites}), we can compute the modified quality factor as (Eq.\,(\ref{Qeffestimation}))
\begin{equation}
Q_2' = (7 \pm 3) \times 10^4 \ .
\end{equation}

For the remainder of this study and also in Paper~II, we adopt $Q_k = 500$ for all satellites.
We then also obtain (Table~\ref{tab:orbital_physical_properties_satellites})
\begin{equation}\label{k2_S1}
    \begin{split}
\frac{k_{2,1}}{Q_1} &= \num{1.77e-6}  &\Leftrightarrow \qquad k_{2,1} \, \dt_1 &= 0.034~\mathrm{s}  \, ,\\
    \frac{k_{2,2}}{Q_2} &= \num{2.04e-5}  &\Leftrightarrow \qquad k_{2,2} \, \dt_2 &= 0.707~\mathrm{s} \ , \\
\frac{k_{2,3}}{Q_3} &= \num{1.47e-5}  &\Leftrightarrow \qquad k_{2,3} \, \dt_3 &= 0.837~\mathrm{s}  \ ,\\
    \frac{k_{2,4}}{Q_4} &= \num{3.98e-5}  &\Leftrightarrow \qquad k_{2,4} \, \dt_4 &= 4.764~\mathrm{s} \ ,\\
    \frac{k_{2,5}}{Q_5} &= \num{3.36e-5}  &\Leftrightarrow \qquad k_{2,5} \, \dt_5 &= 6.219~\mathrm{s} \ ,
    \end{split}
\end{equation} 
which yields  (Eq.\,(\ref{damp:ecc}))
\begin{equation}\label{eq:timescales_eccentricity}
    \begin{split}
	\damp_{e,1} & = 1.25~\mathrm{Gyr} \, ,\\
	\damp_{e,2} & = 0.26~\mathrm{Gyr} \, ,\\
	\damp_{e,3} & = 3.50~\mathrm{Gyr} \, ,\\
	\damp_{e,4} & = 18.7~\mathrm{Gyr} \, ,\\
	\damp_{e,5} & = 155 ~\mathrm{Gyr} \, .
    \end{split}
\end{equation} 
We conclude that tides are able to modify the eccentricities of Miranda, Ariel, and Umbriel over the age of the Solar System, but not those of Titania and Oberon.

\subsection{Inclination damping}\label{sec:inc_damp}

As for the eccentricity, tides always damp the inclination of the satellites (Eq.\,(\ref{eq:inclination_tidal_evolution_full})). 
However, the evolution of the inclination is essentially governed by tides raised on Uranus (Eq.\,(\ref{eq:inclination_tidal_evolution})), as it happens for the semi-major axes (Eq.\,(\ref{eq:semi_major_axis_tidal_evolution})).
The timescale to damp the inclinations is then
\begin{equation}
\label{damp:inc}
    \damp_{\inc,k} \approx \frac{1}{\KtU} \ ,
\end{equation}
which yields about 
\begin{equation}\label{eq:timescales_inclination}
    \begin{split}
	\damp_{\inc,1} & = 290~\mathrm{Gyr} \, ,\\
	\damp_{\inc,2} & = 170~\mathrm{Gyr} \, ,\\
	\damp_{\inc,3} & = 1\,500~\mathrm{Gyr} \, ,\\
	\damp_{\inc,4} & = 14\,300~\mathrm{Gyr} \, ,\\
	\damp_{\inc,5} & = 105\,000~\mathrm{Gyr} \, .
    \end{split}
\end{equation} 
We conclude that tides are very inefficient to damp the inclination of all the satellites.
Indeed, since $ \damp_{\inc,k} \gg \damp_{e,k}$, contrarily to the eccentricity, the present inclinations were likely almost unchanged since the system encountered the 5/3~MMR, and can be seen as fossilised values.


\subsection{Current and damped orbital elements}\label{sec:free_forced_orbital_elements}

In a short time scale, the eccentricity and the inclination of the satellites undergo quasi-periodic variations around an equilibrium state dominated by the secular modes (Tables~\ref{tab:secular_modes_ecc} and~\ref{tab:secular_modes_inc}).
However, as a result of tidal friction, the amplitude of the secular proper modes is damped, and the system evolves into a more quiet state \citep[e.g.][]{Mardling_2007, Laskar_etal_2012}.
In particular, the damping timescale of Ariel's eccentricity is around 260~Myr (Eq.\,(\ref{eq:timescales_eccentricity})), and so we would expect that the amplitude of its proper mode 
to be fully damped since the formation of the system.

To estimate the current variations in the orbital elements, we disabled tidal effects and integrate the present configuration of the Uranian system (Table~\ref{tab:elliptical_elements_Uranus_equatorial_frame}) over $50\,000$ years.
Furthermore, in addition to the five regular satellites, we performed one integration which also included the Sun and another without it. 
The results are presented in Table~\ref{tab:free_eccentricities_inclinations}. 
We note that there are no significant differences with and without the gravitational effect from the Sun, despite the high obliquity of Uranus.
Such happens because Uranus is one of the outermost planets of the Solar System, and perturbations from the Sun are three orders of magnitude smaller than those from the satellites. 
These results confirm that the Uranian system can thus be analysed as an isolated system within our Solar System.

\begin{table*}
    \centering
    \caption{Current variations in the orbital elements of the regular Uranus satellites, with and without the influence of the Sun.}
    \label{tab:free_eccentricities_inclinations}
    \begin{tabular}{c|c|c|c||c|c|c}
        \hline
         \multicolumn{7}{c}{Without the Sun}  \\
        \hline
        Satellite  & $e_{min}$ & $e_{mean}$ & $e_{max}$ & $\inc_{min} \, (\degree)$ & $\inc_{mean} \, (\degree)$ & $\inc_{max} \, (\degree)$\\
         \hline
        Miranda & $\num{8.62e-4}$ & $\num{1.31e-3}$ &$\num{1.76e-3}$ &
        $\num{4.404}$ & $\num{4.409}$ &$\num{4.414}$  \\
        Ariel & $\num{5.77e-5}$ & $\num{1.29e-3}$ &$\num{2.25e-3}$ & 
        $\num{7.92e-4}$ & $\num{2.51e-2}$ &$\num{5.90e-2}$ \\
        Umbriel & $\num{2.20e-3}$ & $\num{3.67e-3}$ &$\num{5.06e-3}$ & 
        $\num{2.41e-3}$ & $\num{7.55e-2}$ &$\num{1.36e-1}$  \\
        Titania & $\num{2.43e-5}$ & $\num{1.64e-3}$ &$\num{3.48e-3}$ & 
        $\num{2.48e-2}$ & $\num{1.22e-1}$ &$\num{1.93e-1}$  \\
        Oberon & $\num{3.64e-5}$ & $\num{1.75e-3}$ &$\num{3.46e-3}$ & 
        $\num{6.62e-2}$ & $\num{1.39e-1}$ &$\num{1.96e-1}$  \\
         \hline \hline
         \multicolumn{7}{c}{With the Sun} \\
        \hline
        Miranda & $\num{8.44e-4}$ & $\num{1.31e-3}$ & $\num{1.76e-3}$ & 
        $\num{4.373}$ & $\num{4.409}$ & $\num{4.445}$  \\
        Ariel & $\num{5.27e-5}$ & $\num{1.29e-3}$ & $\num{2.31e-3}$ & 
        $\num{8.70e-5}$ & $\num{2.72e-2}$ & $\num{8.50e-2}$  \\
        Umbriel & $\num{2.21e-3}$ & $\num{3.53e-3}$ & $\num{5.18e-3}$ & 
        $\num{5.09e-4}$ & $\num{7.35e-2}$ & $\num{1.67e-1}$  \\
        Titania & $\num{1.95e-5}$ & $\num{1.64e-3}$ & $\num{3.50e-3}$ & 
        $\num{1.58e-2}$ & $\num{1.32e-1}$ & $\num{2.59e-1}$  \\
        Oberon & $\num{2.96e-5}$ & $\num{1.82e-3}$ & $\num{3.44e-3}$ & 
        $\num{4.55e-2}$ & $\num{1.72e-1}$ & $\num{2.74e-1}$  \\
        \hline
    \end{tabular}
\end{table*}

To evaluate the expected damped orbital elements, which result from the continuous action of tides, we integrated again the system departing from the current configuration (Table~\ref{tab:elliptical_elements_Uranus_equatorial_frame}).
However, we now consider the tidal dissipation within the satellites, that is, $\dt_k \neq 0$. 
We do not include the tidal dissipation within Uranus ($\dt_0=0$), such that the semi-major axes do not increase (Eq.\,(\ref{eq:semi_major_axis_tidal_evolution})). 
This approximation is justified because the evolution of the eccentricity is dominated by tides raised by Uranus on the satellites (Eq.\,(\ref{eq:eccentricitu_tidal_evolution_simple})).

To ensure that the amplitudes of the proper modes are completely damped, we need to integrate the system over a very long time span, which is computationally expensive.
One solution to decrease the computation time is to enhance the tidal strength by artificially increasing $\dt_k$ and then rescale the evolution time by the same factor. 
This approximation is valid as long as the tidal evolution is adiabatic and also possible because the tidal evolution is proportional to $\dt_k$ (Eq.\,(\ref{tidal_coeficients})).

\begin{figure}
    \centering
    \includegraphics[width=0.5\linewidth]{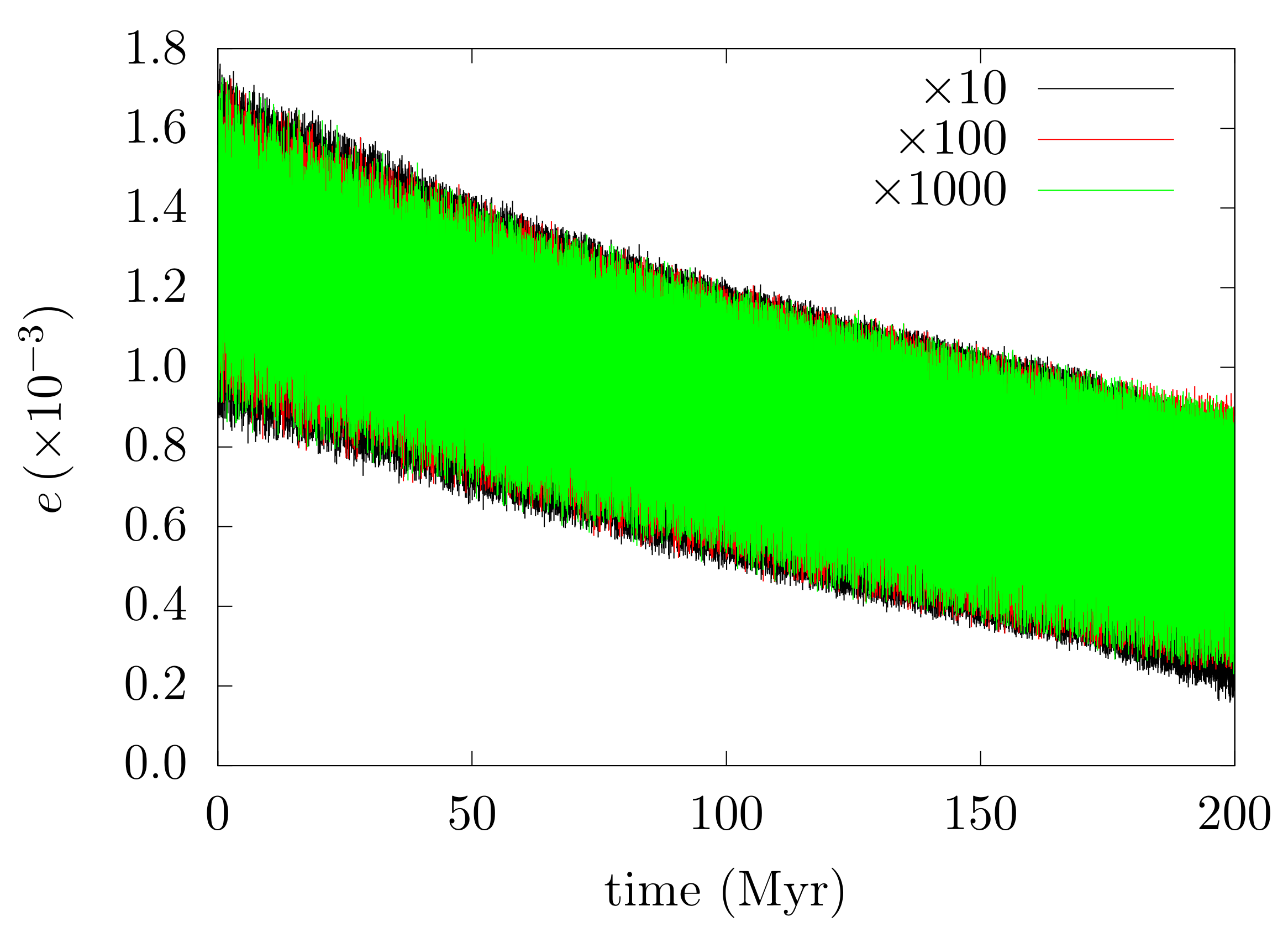}
    \caption{Evolution of the eccentricity of Miranda with three distinct time lags $\dt_1^\prime$: $\dt_1^\prime/\dt_1 = 10$ (black), $\dt_1^\prime/\dt_1=100$ (red), and ${\dt_1^\prime/\dt_1=1000}$ (green).} 
    \label{fig:tidal_multipllication_factors}
\end{figure}

We start by comparing the results from different tidal multiplication factors, $\times 10$, $\times 100$, and $\times 1000$. 
In Fig.\,\ref{fig:tidal_multipllication_factors}, we superimposed the eccentricity of Miranda resulting from an integration over 20 Myr, 2 Myr and 0.2 Myr, where, for each simulation, we rescaled the time axis by the respective multiplication factor. 
We observe that the general evolution of the orbital elements is not sensitive to the enhancement of the tidal strength. 
Therefore, we conclude that we can speed up at least $\times 1000$ the computation time without compromising the results. 
We note that it is not feasible to determine the damped inclinations through this method because we assume $\dt_0=0$ (Sect.~\ref{sec:inc_damp}).
Anyway, the damping timescale of the inclinations is several orders of magnitude larger than the age of the Solar System (Eq.\,(\ref{eq:timescales_inclination})), and so we only need to focus our analysis on the eccentricities.

\begin{figure}
    \centering
    \includegraphics[width=0.5\linewidth]{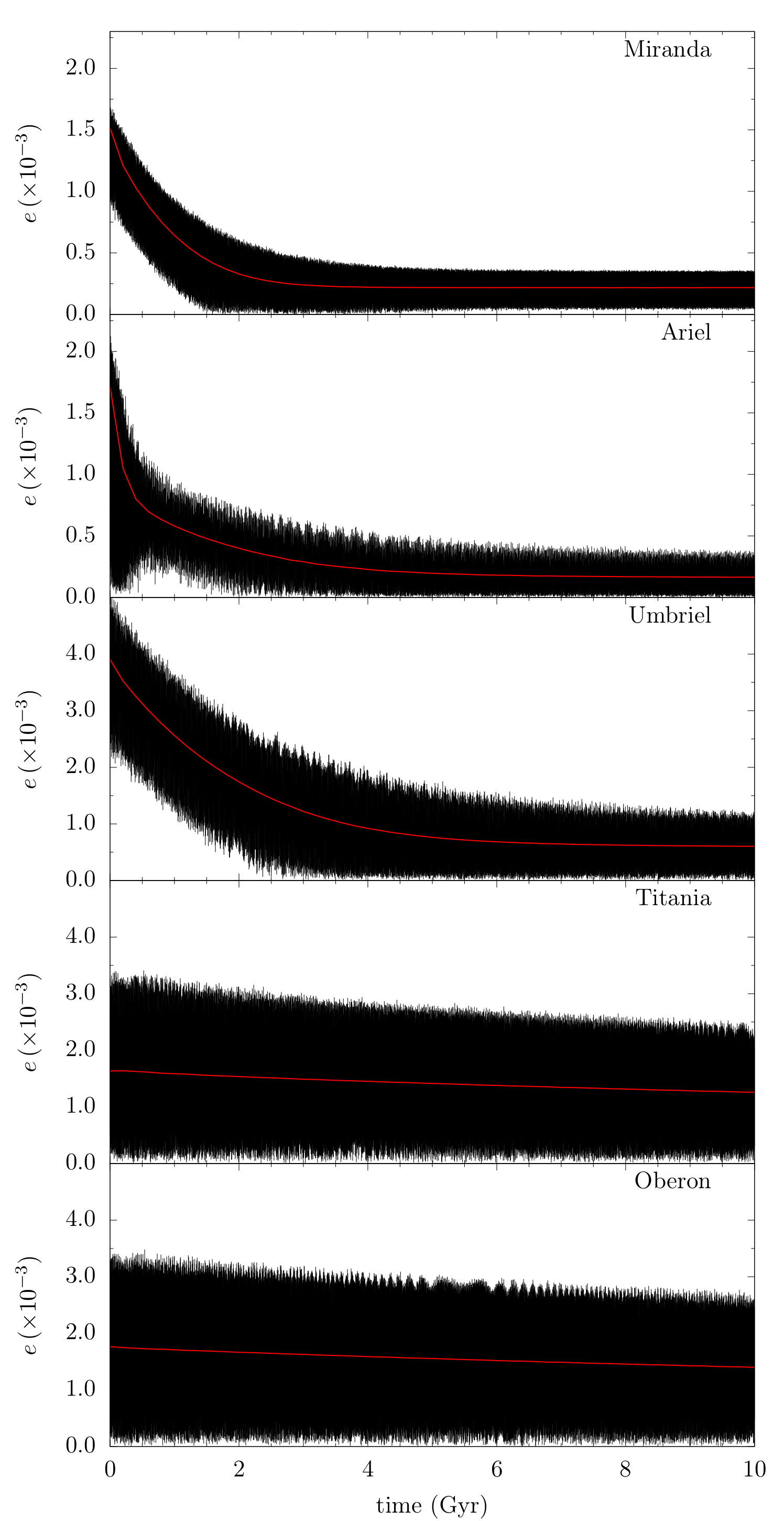}
        \caption{Tidal evolution of the eccentricities of the major Uranian satellites over 10 Gyr. The red line gives the average value of the eccentricity.}
    \label{fig:forced_elements}
\end{figure}

With a multiplication factor of $\times 1000$, we integrated the system over 10~Gyr departing from the current orbital configuration (Table \ref{tab:elliptical_elements_Uranus_equatorial_frame}).
The evolution of the eccentricity for all satellites is shown in Fig. \ref{fig:forced_elements}.
In Table~\ref{tab:forced_orbital_elements}, we list the numerical estimation of the damping timescale for all satellites.
We observe that the eccentricities of Miranda, Ariel, and Umbriel are damped in less than 3~Gyr, which is in good agreement with the theoretical estimations obtained with a single satellite (Eq.\,(\ref{eq:timescales_eccentricity})).
For Titania and Oberon, the decrease in eccentricity over 10~Gyr is small when compared with the other innermost moons, and so we cannot determine neither the damping timescale nor the mean damped eccentricities, we can only put upper limits.
In Table~\ref{tab:forced_orbital_elements}, we provide the mean values of the eccentricity obtained during the last Gyr of evolution.
We observe that for Miranda, Ariel, and Umbriel, the mean damped eccentricity has an amplitude that is about one order of magnitude smaller than the currently observed mean values (Table~\ref{tab:free_eccentricities_inclinations}).
As the eccentricities are still being damped at present, we conclude that some mechanism must have excited the current eccentricities of the three innermost Uranian satellites in a not so distant past (less than 1~Gyr).

\begin{table}
    \centering
    \caption{Mean damped eccentricities for the major Uranian satellites. The values were numerically obtained by integrating the current system for 10~Gyr and disregarding tidal effects from Uranus (Fig.~\ref{fig:forced_elements}).}
    \label{tab:forced_orbital_elements}
    \begin{tabular}{c|c|c}
    	\hline
         Satellite & $ e_{mean}$ & $\damp_{e}$ \\
         \hline
        	Miranda	& $\num{2.1e-4}$ 		& $1.3$ Gyr\\
        	Ariel   & $\num{1.5e-4}$ 		& $0.7$ Gyr\\
        	Umbriel & $\num{6.0e-4}$ 		& $2.9$ Gyr\\
        	Titania & $<\num{1.3e-3}$ 		& $>10$ Gyr\\
        	Oberon  & $<\num{1.3e-3}$ 		& $>10$ Gyr\\
        \hline
    \end{tabular}
\end{table}


\section{Past evolution}\label{sec:backwards_evolution}

The backwards tidal evolution of the semi-major axes obtained in Sect.~\ref{outmigr} disregards the mutual perturbations of the satellites. 
As long as no mean motion resonances are encountered, we expect that the evolution of the complete system does not differ much from the asymptotical evolution provided by Eq.\,(\ref{eq:semi_major_axis_tidal_evolution}).
The same reasoning is also valid for the evolution of the eccentricities (Eq.\,(\ref{eq:eccentricitu_tidal_evolution})) and inclinations (Eq.\,(\ref{eq:inclination_tidal_evolution})).

The closest past lower order mean motion resonance is the 5/3~MMR between Ariel and Umbriel (Fig.~\ref{fig:MMR_crossing}).
The crossing of this resonance shifts the semi-major axes from their asymptotic values and greatly excites the eccentricities (see Paper~II). 
As a consequence, we cannot extend beyond that point the backwards analysis from Sect.~\ref{outmigr}.
However, three-body resonances may also impact the orbital evolution of the system in a near past and set the semi-major axis on a different track \citep[e.g.][]{Greenberg_1975, Cuk_etal_2020}.
Indeed, at present the orbit of Miranda is perturbed by a three-body Laplace resonance with Ariel and Umbriel, with the argument $\lambda_1 - 3 \lambda_2 + 2 \lambda_3$ \citep{Greenberg_1975, Jacobson_2014}.
Therefore, in order to remount to the past evolution of the system, we cannot rely on the asymptotic evolution (Sect.~\ref{outmigr}) and need to take into account the complete $N-$body model described in Sect.~\ref{sec:num_code}.

The resonance crossing is a stochastic and irreversible process, and so we cannot just reverse the time and integrate the system into the past.
We need to place the system where we think it was and then integrate it forwards aiming to reproduce the present observations.
The crossing of the 5/3~MMR is studied in detail in Paper~II.
Here, we follow the evolution from just after the passage through the 5/3~MMR until the present day.
In order to guess the initial conditions, we adopt the asymptotic equations for the semi-major axes (Eq.\,(\ref{eq:semi_major_axis_tidal_evolution})) and eccentricities (Eq.\,(\ref{eq:eccentricitu_tidal_evolution})) without coupling, leading to
\begin{equation}
    \begin{split}
      & a_1 / \Ru = 5.0584  \ , \quad  & e_1= \num{2.7e-03} \ , \\
      & a_2 / \Ru = 7.3959  \ , \quad  & e_2= \num{8.0e-03} \ ,\\
      & a_3 / \Ru = 10.3900  \ , \quad  & e_3= \num{4.5e-03}  \ ,\\
      & a_4 / \Ru = 17.0645  \ , \quad  & e_4= \num{2.0e-3} \ ,\\
      & a_5 / \Ru = 22.8234  \ , \quad  & e_5= \num{2.0e-3} \ ,
    \end{split}
\end{equation}

For the inclinations, we adopt the present values (Table \ref{tab:orbital_physical_properties_satellites}), as they are not expected to vary much (Sect.~\ref{sec:inc_damp}), except for Miranda's inclination, which was slightly increased to $4.41 \degree$ to match the current inclination at the end of the simulation.
The initial rotations of the satellites are synchronous with the orbital mean motion ($\angrot_k / n_k = 1$), and the initial obliquities were set to zero ($\obq_k=0\degree$).
We then performed 100 simulations, where, for each run, we randomly choose the remaining orbital parameters ($\lambda_k, \varpi_k, \Omega_h$) between $0^\circ$ and $360^\circ$. 
For the dissipation, we adopt the values given by expressions (\ref{k2sQ}) and (\ref{k2_S1}).

\begin{sidewaysfigure*}[htbp]
    \centering
    \includegraphics[width=\textwidth]{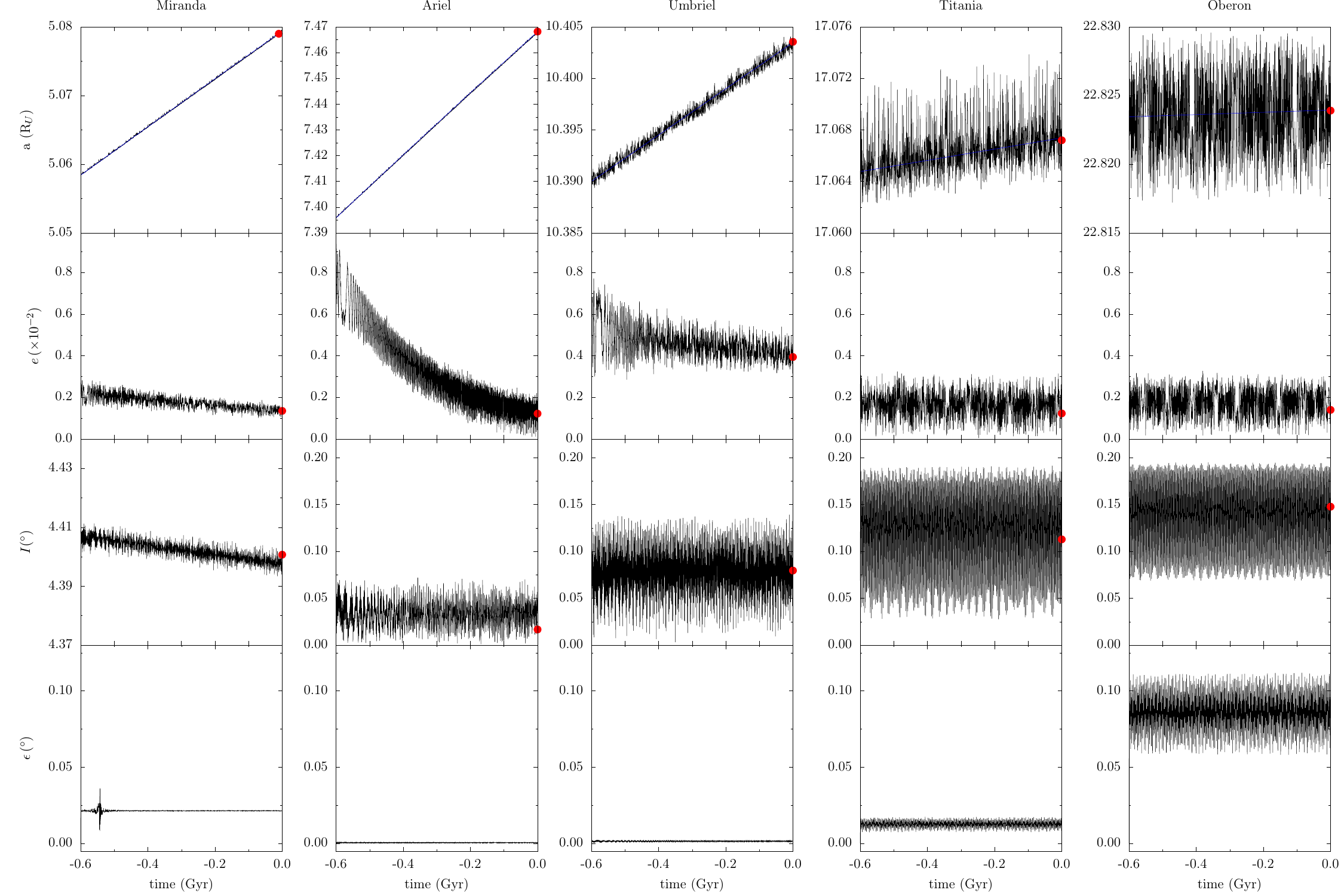}
    \caption{Orbital and spin evolution of the five main satellites of Uranus from 600~Myr ago to the present. From the top to the bottom, we show the semi-major axes, the eccentricities, the inclinations, and the obliquities. The blue line gives the asymptotic evolution of the semi-major axes, obtained by Eq.\,(\ref{eq:semi_major_axis_tidal_evolution}). The current mean orbital parameters (Table~\ref{tab:free_eccentricities_inclinations}) were superimposed as red circles.}
    \label{fig:step_one}
\end{sidewaysfigure*}

In Fig. \ref{fig:step_one}, we display one example of the evolution of the semi-major axes, eccentricities, inclinations, and obliquities, resulting from the integration of these initial conditions. 
As expected, we confirm that no two-body MMR was crossed. 
More importantly, we observe that the system is not captured in any three-body mean motion resonance either, and the migration rates are in excellent agreement with the asymptotic predictions (Eq.\,(\ref{eq:semi_major_axis_tidal_evolution})). 
The final eccentricities and inclinations are also in perfect agreement with the currently observed values (Table~\ref{tab:orbital_physical_properties_satellites}).

Similar results were observed in all the 100 simulations that we performed.
No unexpected event deviated the system from its course, and all the 100 simulations correctly reproduced the current system.
Our simulations cannot completetly rule out a different evolution scenario, but it should certainly have a low probability.
Therefore, we conclude that, after the passage through the 5/3 MMR between Ariel and Umbriel, the evolution of the system is mostly peaceful and dominated by tides raised on Uranus by the satellites.

In our simulations, we also observe that, starting from $\obq_k=0\degree$, the obliquities of all satellites  quickly evolve\footnote{Because of the temporal resolution of the graphs, it is not possible to observe the rapid initial evolution of the obliquities, which may give the impression that the initial obliquities adopted were not zero.} into the low obliquity Cassini state~1 (see Table~\ref{tab:cassini_state_1}), corroborating that 
it is unlikely for the satellites to become involved in the resonant high obliquity Cassini state~2, as suggested by \citet{Cuk_etal_2020}.
Indeed, we have conducted some additional simulations with initial obliquities spanning between $0\degree$ and $180\degree$, and observed that the obliquity always evolves into Cassini State~1, indicating that the initial obliquity choice is not critical.


\section{Summary and Discussion}\label{sec:conclusion}

In the short-term, the variations in the orbits of Uranus's largest satellites are governed by mutual perturbations.
However, the past long-term evolution is driven by tidal effects between the planet and its satellites.
In this work, we have studied the present dynamics and the recent tidal evolution of these moons, namely, Miranda, Ariel, Umbriel, Titania, and Oberon.
For that purpose, we have used a $N-$body code that takes into account satellite-satellite interactions, spin dynamics, and tidal dissipation. 

We begun our study by accessing the present orbital configuration of the system.
Departing from more recent data given by \citet{Jacobson_2014}, we recomputed the secular modes of the five main satellites using the frequency analysis method \citep[as in][]{Laskar_1986, Laskar_Jacobson_1987}. 
With these updated values, we accessed the possibility of spin-orbit resonances within the major satellites of Uranus.
\citet{Cuk_etal_2020} suggested that the current high inclination of Miranda could be explained due to the crossing of the $5/3$~MMR between Ariel and Umbriel.
In this scenario, the inclinations of all secular modes are excited, but their amplitude could then be lowered through spin-orbit resonances.
We show that the only possibility is for a resonance between the precession of the spin axis of Oberon and the $s_3$ secular mode.
Therefore, it is not feasible to damp the amplitude of the remaining secular models through this mechanism, and
a more suitable explanation for the high inclination of Miranda remains the crossing of the $3/1$~MMR between Miranda and Umbriel \citep{Tittemore_Wisdom_1989, Tittemore_Wisdom_1990}.

\citet{Tittemore_Wisdom_1990} argue that the 2/1~MMR between Miranda and Ariel could not have been crossed in the past, because capture is certain and the system cannot evade it. 
On the other hand, the system has to cross the $3/1$~MMR between Miranda and Umbriel to increase the inclination of Miranda.
Using these constraints, we updated the quality factor of Uranus to be ${5\,800<Q_0<11\,500}$, which extends beyond the previously established range outlined by \citet{Tittemore_Wisdom_1990}. 
This is a consequence of the improvement in the determination of the physical parameters of the satellites, namely, in the radii and the masses \citep{Jacobson_2014}.
Another consequence is that the encounter with the 5/3~MMR between Ariel and Umbriel must have occurred about 460 to 920~Myr ago.

The crossing of the 5/3~MMR excites the eccentricities of all satellites, but we show that tides quickly erode those of Miranda, Ariel, and Umbriel.
In order to comply with the currently observed eccentricity values, we estimate that the quality factor of Ariel is $Q_2 \sim 500$.
The inclinations remain essentially unchanged and can therefore be used as a fossilised signature of the system that emerged after the 5/3~MMR.

By performing a large number of numerical simulations, we finally show that, from the passage through the 5/3~MMR to the present, the system is most likely not captured in any three-body mean motion resonance.
As a result, the orbital evolution of each satellite follows the tidal evolution derived from an unperturbed two-body problem.
In the companion Paper~II \citep{Gomes_Correia_2024p2}, we provide an exhaustive study on the crossing of the 5/3~MMR that sticks with the analysis presented here.


\section*{Acknowledgements}
This work was supported by COMPETE 2020 and by
FCT - Funda\c{c}\~ao para a Ci\^encia e a Tecnologia, I.P., Portugal, 
through the projects
SFRH/BD/143371/2019,
GRAVITY (PTDC/FIS-AST/7002/2020),
ENGAGE SKA (POCI-01-0145-FEDER-022217), and
CFisUC (UIDB/04564/2020 and UIDP/04564/2020, with DOI identifiers 10.54499/UIDB/04564/2020 and 10.54499/UIDP/04564/2020, respectively).
We acknowledge the Laboratory for Advanced Computing at University of Coimbra (\href{https://www.uc.pt/lca}{https://www.uc.pt/lca}) for providing the resources to perform the numerical simulations.


\bibliographystyle{cas-model2-names}

\bibliography{bibliography.bib}

\newcommand{\noop}[1]{}
\begin{thebibliography}{77}
\expandafter\ifx\csname natexlab\endcsname\relax\def\natexlab#1{#1}\fi
\providecommand{\url}[1]{\texttt{#1}}
\providecommand{\href}[2]{#2}
\providecommand{\path}[1]{#1}
\providecommand{\DOIprefix}{doi:}
\providecommand{\ArXivprefix}{arXiv:}
\providecommand{\URLprefix}{URL: }
\providecommand{\Pubmedprefix}{pmid:}
\providecommand{\doi}[1]{\href{http://dx.doi.org/#1}{\path{#1}}}
\providecommand{\Pubmed}[1]{\href{pmid:#1}{\path{#1}}}
\providecommand{\bibinfo}[2]{#2}
\ifx\xfnm\relax \def\xfnm[#1]{\unskip,\space#1}\fi
\bibitem[{{Agnor} and {Hamilton}(2006)}]{Agnor_2006}
\bibinfo{author}{{Agnor}, C.B.}, \bibinfo{author}{{Hamilton}, D.P.},
  \bibinfo{year}{2006}.
\newblock \bibinfo{title}{{Neptune's capture of its moon Triton in a
  binary-planet gravitational encounter}}.
\newblock \bibinfo{journal}{\nat} \bibinfo{volume}{441},
  \bibinfo{pages}{192--194}.
\newblock \DOIprefix\doi{10.1038/nature04792}.
\bibitem[{{Alexander}(1973)}]{Alexander_1973}
\bibinfo{author}{{Alexander}, M.E.}, \bibinfo{year}{1973}.
\newblock \bibinfo{title}{{The Weak Friction Approximation and Tidal Evolution
  in Close Binary Systems}}.
\newblock \bibinfo{journal}{\apss} \bibinfo{volume}{23},
  \bibinfo{pages}{459--510}.
\newblock \DOIprefix\doi{10.1007/BF00645172}.
\bibitem[{{Avramchuk} et~al.(2007){Avramchuk}, {Rosenbush} and
  {Bul'Ba}}]{Avramchuk_etal_2007}
\bibinfo{author}{{Avramchuk}, V.V.}, \bibinfo{author}{{Rosenbush}, V.K.},
  \bibinfo{author}{{Bul'Ba}, T.P.}, \bibinfo{year}{2007}.
\newblock \bibinfo{title}{{Photometric study of the major satellites of
  Uranus}}.
\newblock \bibinfo{journal}{Solar System Research} \bibinfo{volume}{41},
  \bibinfo{pages}{186--202}.
\newblock \DOIprefix\doi{10.1134/S0038094607030021}.
\bibitem[{{Beddingfield} et~al.(2022){Beddingfield}, {Cartwright}, {Leonard},
  {Nordheim} and {Scipioni}}]{Beddingfield_etal_2022}
\bibinfo{author}{{Beddingfield}, C.B.}, \bibinfo{author}{{Cartwright}, R.J.},
  \bibinfo{author}{{Leonard}, E.}, \bibinfo{author}{{Nordheim}, T.},
  \bibinfo{author}{{Scipioni}, F.}, \bibinfo{year}{2022}.
\newblock \bibinfo{title}{{Ariel's Elastic Thicknesses and Heat Fluxes}}.
\newblock \bibinfo{journal}{\psj} \bibinfo{volume}{3}, \bibinfo{pages}{106}.
\newblock \DOIprefix\doi{10.3847/PSJ/ac63d1}.
\bibitem[{{Beddingfield} et~al.(2023){Beddingfield}, {Leonard}, {Nordheim},
  {Cartwright} and {Castillo-Rogez}}]{Beddingfield_etal_2023}
\bibinfo{author}{{Beddingfield}, C.B.}, \bibinfo{author}{{Leonard}, E.J.},
  \bibinfo{author}{{Nordheim}, T.A.}, \bibinfo{author}{{Cartwright}, R.J.},
  \bibinfo{author}{{Castillo-Rogez}, J.C.}, \bibinfo{year}{2023}.
\newblock \bibinfo{title}{{Titania's Heat Fluxes Revealed by Messina
  Chasmata}}.
\newblock \bibinfo{journal}{\psj} \bibinfo{volume}{4}, \bibinfo{pages}{211}.
\newblock \DOIprefix\doi{10.3847/PSJ/ad0367}.
\bibitem[{{Boekholt} and {Correia}(2023)}]{Boekholt_Correia_2023}
\bibinfo{author}{{Boekholt}, T.C.N.}, \bibinfo{author}{{Correia}, A.C.M.},
  \bibinfo{year}{2023}.
\newblock \bibinfo{title}{{A direct N-body integrator for modelling the
  chaotic, tidal dynamics of multibody extrasolar systems: TIDYMESS}}.
\newblock \bibinfo{journal}{\mnras} \bibinfo{volume}{522},
  \bibinfo{pages}{2885--2900}.
\newblock \DOIprefix\doi{10.1093/mnras/stad1133},
  \href{http://arxiv.org/abs/2209.03955}{\tt arXiv:2209.03955}.
\bibitem[{{Bottke} et~al.(2024){Bottke}, {Vokrouhlick{\'y}}, {Nesvorn{\'y}},
  {Marshall}, {Morbidelli}, {Deienno}, {Marchi}, {Kirchoff}, {Dones} and
  {Levison}}]{Bottke_etal_2024}
\bibinfo{author}{{Bottke}, W.F.}, \bibinfo{author}{{Vokrouhlick{\'y}}, D.},
  \bibinfo{author}{{Nesvorn{\'y}}, D.}, \bibinfo{author}{{Marshall}, R.},
  \bibinfo{author}{{Morbidelli}, A.}, \bibinfo{author}{{Deienno}, R.},
  \bibinfo{author}{{Marchi}, S.}, \bibinfo{author}{{Kirchoff}, M.},
  \bibinfo{author}{{Dones}, L.}, \bibinfo{author}{{Levison}, H.F.},
  \bibinfo{year}{2024}.
\newblock \bibinfo{title}{{The Bombardment History of the Giant Planet
  Satellites}}.
\newblock \bibinfo{journal}{\psj} \bibinfo{volume}{5}, \bibinfo{pages}{88}.
\newblock \DOIprefix\doi{10.3847/PSJ/ad29f4}.
\bibitem[{{Bou{\'e}} and {Laskar}(2010)}]{Boue_Laskar_2010}
\bibinfo{author}{{Bou{\'e}}, G.}, \bibinfo{author}{{Laskar}, J.},
  \bibinfo{year}{2010}.
\newblock \bibinfo{title}{{A Collisionless Scenario for Uranus Tilting}}.
\newblock \bibinfo{journal}{The Astrophysical Journal Letters}
  \bibinfo{volume}{712}, \bibinfo{pages}{L44--L47}.
\newblock \DOIprefix\doi{10.1088/2041-8205/712/1/L44},
  \href{http://arxiv.org/abs/0912.0181}{\tt arXiv:0912.0181}.
\bibitem[{{Carpino} et~al.(1987){Carpino}, {Milani} and
  {Nobili}}]{Carpino_etal_1987}
\bibinfo{author}{{Carpino}, M.}, \bibinfo{author}{{Milani}, A.},
  \bibinfo{author}{{Nobili}, A.M.}, \bibinfo{year}{1987}.
\newblock \bibinfo{title}{{Long-term numerical integrations and synthetic
  theories for the motion of the outer planets}}.
\newblock \bibinfo{journal}{\aap} \bibinfo{volume}{181},
  \bibinfo{pages}{182--194}.
\bibitem[{{Chen} et~al.(2014){Chen}, {Nimmo} and {Glatzmaier}}]{Chen_etal_2014}
\bibinfo{author}{{Chen}, E.M.A.}, \bibinfo{author}{{Nimmo}, F.},
  \bibinfo{author}{{Glatzmaier}, G.A.}, \bibinfo{year}{2014}.
\newblock \bibinfo{title}{{Tidal heating in icy satellite oceans}}.
\newblock \bibinfo{journal}{\icarus} \bibinfo{volume}{229},
  \bibinfo{pages}{11--30}.
\newblock \DOIprefix\doi{10.1016/j.icarus.2013.10.024}.
\bibitem[{{Colombo}(1966)}]{Colombo_1966}
\bibinfo{author}{{Colombo}, G.}, \bibinfo{year}{1966}.
\newblock \bibinfo{title}{{Cassini's second and third laws}}.
\newblock \bibinfo{journal}{\aj} \bibinfo{volume}{71},
  \bibinfo{pages}{891--896}.
\bibitem[{{Correia}(2009)}]{Correia_2009}
\bibinfo{author}{{Correia}, A.C.M.}, \bibinfo{year}{2009}.
\newblock \bibinfo{title}{{Secular Evolution of a Satellite by Tidal Effect:
  Application to Triton}}.
\newblock \bibinfo{journal}{\apjl} \bibinfo{volume}{704},
  \bibinfo{pages}{L1--L4}.
\newblock \DOIprefix\doi{10.1088/0004-637X/704/1/L1},
  \href{http://arxiv.org/abs/0909.4210}{\tt arXiv:0909.4210}.
\bibitem[{{Correia}(2015)}]{Correia_2015}
\bibinfo{author}{{Correia}, A.C.M.}, \bibinfo{year}{2015}.
\newblock \bibinfo{title}{{Stellar and planetary Cassini states}}.
\newblock \bibinfo{journal}{\aap} \bibinfo{volume}{582}, \bibinfo{pages}{A69}.
\newblock \DOIprefix\doi{10.1051/0004-6361/201525939}.
\bibitem[{{Correia}(2018)}]{Correia_2018}
\bibinfo{author}{{Correia}, A.C.M.}, \bibinfo{year}{2018}.
\newblock \bibinfo{title}{{Chaotic dynamics in the (47171) Lempo triple
  system}}.
\newblock \bibinfo{journal}{\icarus} \bibinfo{volume}{305},
  \bibinfo{pages}{250--261}.
\newblock \DOIprefix\doi{10.1016/j.icarus.2018.01.008},
  \href{http://arxiv.org/abs/1710.08401}{\tt arXiv:1710.08401}.
\bibitem[{{Correia} et~al.(2016){Correia}, {Bou{\'e}} and
  {Laskar}}]{Correia_etal_2016}
\bibinfo{author}{{Correia}, A.C.M.}, \bibinfo{author}{{Bou{\'e}}, G.},
  \bibinfo{author}{{Laskar}, J.}, \bibinfo{year}{2016}.
\newblock \bibinfo{title}{{Secular and tidal evolution of circumbinary
  systems}}.
\newblock \bibinfo{journal}{Celestial Mechanics and Dynamical Astronomy}
  \bibinfo{volume}{126}, \bibinfo{pages}{189--225}.
\newblock \DOIprefix\doi{10.1007/s10569-016-9709-9},
  \href{http://arxiv.org/abs/1608.03484}{\tt arXiv:1608.03484}.
\bibitem[{{Correia} and {Laskar}(2003)}]{Correia_Laskar_2003I}
\bibinfo{author}{{Correia}, A.C.M.}, \bibinfo{author}{{Laskar}, J.},
  \bibinfo{year}{2003}.
\newblock \bibinfo{title}{{Long-term evolution of the spin of Venus II.
  Numerical simulations}}.
\newblock \bibinfo{journal}{\icarus} \bibinfo{volume}{163},
  \bibinfo{pages}{24--45}.
\newblock \DOIprefix\doi{10.1016/S0019-1035(03)00043-5}.
\bibitem[{{Correia} et~al.(2011){Correia}, {Laskar}, {Farago} and
  {Bou{\'e}}}]{Correia_etal_2011}
\bibinfo{author}{{Correia}, A.C.M.}, \bibinfo{author}{{Laskar}, J.},
  \bibinfo{author}{{Farago}, F.}, \bibinfo{author}{{Bou{\'e}}, G.},
  \bibinfo{year}{2011}.
\newblock \bibinfo{title}{{Tidal evolution of hierarchical and inclined
  systems}}.
\newblock \bibinfo{journal}{Celestial Mechanics and Dynamical Astronomy}
  \bibinfo{volume}{111}, \bibinfo{pages}{105--130}.
\newblock \DOIprefix\doi{10.1007/s10569-011-9368-9},
  \href{http://arxiv.org/abs/1107.0736}{\tt arXiv:1107.0736}.
\bibitem[{{Correia} and {Rodr{\'\i}guez}(2013)}]{Correia_Rodriguez_2013}
\bibinfo{author}{{Correia}, A.C.M.}, \bibinfo{author}{{Rodr{\'\i}guez}, A.},
  \bibinfo{year}{2013}.
\newblock \bibinfo{title}{{On the Equilibrium Figure of Close-in Planets and
  Satellites}}.
\newblock \bibinfo{journal}{\apj} \bibinfo{volume}{767}, \bibinfo{pages}{128}.
\newblock \DOIprefix\doi{10.1088/0004-637X/767/2/128},
  \href{http://arxiv.org/abs/1304.1425}{\tt arXiv:1304.1425}.
\bibitem[{{Correia} and {Valente}(2022)}]{Correia_Valente_2022}
\bibinfo{author}{{Correia}, A.C.M.}, \bibinfo{author}{{Valente}, E.F.S.},
  \bibinfo{year}{2022}.
\newblock \bibinfo{title}{{Tidal evolution for any rheological model using a
  vectorial approach expressed in Hansen coefficients}}.
\newblock \bibinfo{journal}{Celestial Mechanics and Dynamical Astronomy}
  \bibinfo{volume}{134}, \bibinfo{pages}{24}.
\newblock \DOIprefix\doi{10.1007/s10569-022-10079-3},
  \href{http://arxiv.org/abs/2306.03449}{\tt arXiv:2306.03449}.
\bibitem[{{{\'C}uk} et~al.(2016){{\'C}uk}, {Dones} and
  {Nesvorn{\'y}}}]{Cuk_etal_2016a}
\bibinfo{author}{{{\'C}uk}, M.}, \bibinfo{author}{{Dones}, L.},
  \bibinfo{author}{{Nesvorn{\'y}}, D.}, \bibinfo{year}{2016}.
\newblock \bibinfo{title}{{Dynamical Evidence for a Late Formation of
  Saturn{\textquoteright}s Moons}}.
\newblock \bibinfo{journal}{\apj} \bibinfo{volume}{820}, \bibinfo{pages}{97}.
\newblock \DOIprefix\doi{10.3847/0004-637X/820/2/97},
  \href{http://arxiv.org/abs/1603.07071}{\tt arXiv:1603.07071}.
\bibitem[{{{\'C}uk} et~al.(2020){{\'C}uk}, {El Moutamid} and
  {Tiscareno}}]{Cuk_etal_2020}
\bibinfo{author}{{{\'C}uk}, M.}, \bibinfo{author}{{El Moutamid}, M.},
  \bibinfo{author}{{Tiscareno}, M.S.}, \bibinfo{year}{2020}.
\newblock \bibinfo{title}{{Dynamical History of the Uranian System}}.
\newblock \bibinfo{journal}{\psj} \bibinfo{volume}{1}, \bibinfo{pages}{22}.
\newblock \DOIprefix\doi{10.3847/PSJ/ab9748},
  \href{http://arxiv.org/abs/2005.12887}{\tt arXiv:2005.12887}.
\bibitem[{{Dai} et~al.(2023){Dai}, {Masuda}, {Beard}, {Robertson}, {Goldberg},
  {Batygin}, {Bouma}, {Lissauer}, {Knudstrup}, {Albrecht}, {Howard}, {Knutson},
  {Petigura}, {Weiss}, {Isaacson}, {Kristiansen}, {Osborn}, {Wang}, {Wang},
  {Behmard}, {Greklek-McKeon}, {Vissapragada}, {Batalha}, {Brinkman},
  {Chontos}, {Crossfield}, {Dressing}, {Fetherolf}, {Fulton}, {Hill}, {Huber},
  {Kane}, {Lubin}, {MacDougall}, {Mayo}, {Mo{\v{c}}nik}, {Akana Murphy},
  {Rubenzahl}, {Scarsdale}, {Tyler}, {Zandt}, {Polanski}, {Schwengeler},
  {Terentev}, {Benni}, {Bieryla}, {Ciardi}, {Falk}, {Furlan}, {Girardin},
  {Guerra}, {Hesse}, {Howell}, {Lillo-Box}, {Matthews}, {Twicken},
  {Villase{\~n}or}, {Latham}, {Jenkins}, {Ricker}, {Seager}, {Vanderspek} and
  {Winn}}]{Dai_etal_2023}
\bibinfo{author}{{Dai}, F.}, \bibinfo{author}{{Masuda}, K.},
  \bibinfo{author}{{Beard}, C.}, \bibinfo{author}{{Robertson}, P.},
  \bibinfo{author}{{Goldberg}, M.}, \bibinfo{author}{{Batygin}, K.},
  \bibinfo{author}{{Bouma}, L.}, \bibinfo{author}{{Lissauer}, J.J.},
  \bibinfo{author}{{Knudstrup}, E.}, \bibinfo{author}{{Albrecht}, S.},
  \bibinfo{author}{{Howard}, A.W.}, \bibinfo{author}{{Knutson}, H.A.},
  \bibinfo{author}{{Petigura}, E.A.}, \bibinfo{author}{{Weiss}, L.M.},
  \bibinfo{author}{{Isaacson}, H.}, \bibinfo{author}{{Kristiansen}, M.H.},
  \bibinfo{author}{{Osborn}, H.}, \bibinfo{author}{{Wang}, S.},
  \bibinfo{author}{{Wang}, X.Y.}, \bibinfo{author}{{Behmard}, A.},
  \bibinfo{author}{{Greklek-McKeon}, M.}, \bibinfo{author}{{Vissapragada}, S.},
  \bibinfo{author}{{Batalha}, N.M.}, \bibinfo{author}{{Brinkman}, C.L.},
  \bibinfo{author}{{Chontos}, A.}, \bibinfo{author}{{Crossfield}, I.},
  \bibinfo{author}{{Dressing}, C.}, \bibinfo{author}{{Fetherolf}, T.},
  \bibinfo{author}{{Fulton}, B.}, \bibinfo{author}{{Hill}, M.L.},
  \bibinfo{author}{{Huber}, D.}, \bibinfo{author}{{Kane}, S.R.},
  \bibinfo{author}{{Lubin}, J.}, \bibinfo{author}{{MacDougall}, M.},
  \bibinfo{author}{{Mayo}, A.}, \bibinfo{author}{{Mo{\v{c}}nik}, T.},
  \bibinfo{author}{{Akana Murphy}, J.M.}, \bibinfo{author}{{Rubenzahl}, R.A.},
  \bibinfo{author}{{Scarsdale}, N.}, \bibinfo{author}{{Tyler}, D.},
  \bibinfo{author}{{Zandt}, J.V.}, \bibinfo{author}{{Polanski}, A.S.},
  \bibinfo{author}{{Schwengeler}, H.M.}, \bibinfo{author}{{Terentev}, I.A.},
  \bibinfo{author}{{Benni}, P.}, \bibinfo{author}{{Bieryla}, A.},
  \bibinfo{author}{{Ciardi}, D.}, \bibinfo{author}{{Falk}, B.},
  \bibinfo{author}{{Furlan}, E.}, \bibinfo{author}{{Girardin}, E.},
  \bibinfo{author}{{Guerra}, P.}, \bibinfo{author}{{Hesse}, K.M.},
  \bibinfo{author}{{Howell}, S.B.}, \bibinfo{author}{{Lillo-Box}, J.},
  \bibinfo{author}{{Matthews}, E.C.}, \bibinfo{author}{{Twicken}, J.D.},
  \bibinfo{author}{{Villase{\~n}or}, J.}, \bibinfo{author}{{Latham}, D.W.},
  \bibinfo{author}{{Jenkins}, J.M.}, \bibinfo{author}{{Ricker}, G.R.},
  \bibinfo{author}{{Seager}, S.}, \bibinfo{author}{{Vanderspek}, R.},
  \bibinfo{author}{{Winn}, J.N.}, \bibinfo{year}{2023}.
\newblock \bibinfo{title}{{TOI-1136 is a Young, Coplanar, Aligned Planetary
  System in a Pristine Resonant Chain}}.
\newblock \bibinfo{journal}{\aj} \bibinfo{volume}{165}, \bibinfo{pages}{33}.
\newblock \DOIprefix\doi{10.3847/1538-3881/aca327},
  \href{http://arxiv.org/abs/2210.09283}{\tt arXiv:2210.09283}.
\bibitem[{{Deienno} et~al.(2011){Deienno}, {Yokoyama}, {Nogueira}, {Callegari}
  and {Santos}}]{Deienno_etal_2011}
\bibinfo{author}{{Deienno}, R.}, \bibinfo{author}{{Yokoyama}, T.},
  \bibinfo{author}{{Nogueira}, E.C.}, \bibinfo{author}{{Callegari}, N.},
  \bibinfo{author}{{Santos}, M.T.}, \bibinfo{year}{2011}.
\newblock \bibinfo{title}{{Effects of the planetary migration on some
  primordial satellites of the outer planets. I. Uranus' case}}.
\newblock \bibinfo{journal}{\aap} \bibinfo{volume}{536}, \bibinfo{pages}{A57}.
\newblock \DOIprefix\doi{10.1051/0004-6361/201014862}.
\bibitem[{{Dermott} et~al.(1988){Dermott}, {Malhotra} and
  {Murray}}]{Dermott_etal_1988}
\bibinfo{author}{{Dermott}, S.F.}, \bibinfo{author}{{Malhotra}, R.},
  \bibinfo{author}{{Murray}, C.D.}, \bibinfo{year}{1988}.
\newblock \bibinfo{title}{{Dynamics of the Uranian and Saturnian satelite
  systems: A chaotic route to melting Miranda?}}
\newblock \bibinfo{journal}{\icarus} \bibinfo{volume}{76},
  \bibinfo{pages}{295--334}.
\newblock \DOIprefix\doi{10.1016/0019-1035(88)90074-7}.
\bibitem[{{Dermott} and {Nicholson}(1986)}]{Dermott_Nicholson_1986}
\bibinfo{author}{{Dermott}, S.F.}, \bibinfo{author}{{Nicholson}, P.D.},
  \bibinfo{year}{1986}.
\newblock \bibinfo{title}{{Masses of the satellites of Uranus}}.
\newblock \bibinfo{journal}{\nat} \bibinfo{volume}{319},
  \bibinfo{pages}{115--120}.
\newblock \DOIprefix\doi{10.1038/319115a0}.
\bibitem[{{Fabrycky} et~al.(2007){Fabrycky}, {Johnson} and
  {Goodman}}]{Fabrycky_etal_2007}
\bibinfo{author}{{Fabrycky}, D.C.}, \bibinfo{author}{{Johnson}, E.T.},
  \bibinfo{author}{{Goodman}, J.}, \bibinfo{year}{2007}.
\newblock \bibinfo{title}{{Cassini States with Dissipation: Why Obliquity Tides
  Cannot Inflate Hot Jupiters}}.
\newblock \bibinfo{journal}{\apj} \bibinfo{volume}{665},
  \bibinfo{pages}{754--766}.
\newblock \DOIprefix\doi{10.1086/519075},
  \href{http://arxiv.org/abs/astro-ph/0703418}{\tt arXiv:astro-ph/0703418}.
\bibitem[{Gastineau and Laskar(2011)}]{TRIP}
\bibinfo{author}{Gastineau, M.}, \bibinfo{author}{Laskar, J.},
  \bibinfo{year}{2011}.
\newblock \bibinfo{title}{Trip: A computer algebra system dedicated to
  celestial mechanics and perturbation series}.
\newblock \bibinfo{journal}{ACM Commun. Comput. Algebra} \bibinfo{volume}{44},
  \bibinfo{pages}{194--197}.
\newblock \URLprefix \url{http://doi.acm.org/10.1145/1940475.1940518},
  \DOIprefix\doi{10.1145/1940475.1940518}.
\bibitem[{{Gavrilov} and {Zharkov}(1977)}]{Gavrilov_Zharkov_1977}
\bibinfo{author}{{Gavrilov}, S.V.}, \bibinfo{author}{{Zharkov}, V.N.},
  \bibinfo{year}{1977}.
\newblock \bibinfo{title}{{Love Numbers of the Giant Planets}}.
\newblock \bibinfo{journal}{\icarus} \bibinfo{volume}{32},
  \bibinfo{pages}{443--449}.
\newblock \DOIprefix\doi{10.1016/0019-1035(77)90015-X}.
\bibitem[{{Gillon} et~al.(2017){Gillon}, {Triaud}, {Demory}, {Jehin}, {Agol},
  {Deck}, {Lederer}, {de Wit}, {Burdanov}, {Ingalls}, {Bolmont}, {Leconte},
  {Raymond}, {Selsis}, {Turbet}, {Barkaoui}, {Burgasser}, {Burleigh}, {Carey},
  {Chaushev}, {Copperwheat}, {Delrez}, {Fernandes}, {Holdsworth}, {Kotze}, {Van
  Grootel}, {Almleaky}, {Benkhaldoun}, {Magain} and
  {Queloz}}]{Gillon_etal_2017}
\bibinfo{author}{{Gillon}, M.}, \bibinfo{author}{{Triaud}, A.H.M.J.},
  \bibinfo{author}{{Demory}, B.O.}, \bibinfo{author}{{Jehin}, E.},
  \bibinfo{author}{{Agol}, E.}, \bibinfo{author}{{Deck}, K.M.},
  \bibinfo{author}{{Lederer}, S.M.}, \bibinfo{author}{{de Wit}, J.},
  \bibinfo{author}{{Burdanov}, A.}, \bibinfo{author}{{Ingalls}, J.G.},
  \bibinfo{author}{{Bolmont}, E.}, \bibinfo{author}{{Leconte}, J.},
  \bibinfo{author}{{Raymond}, S.N.}, \bibinfo{author}{{Selsis}, F.},
  \bibinfo{author}{{Turbet}, M.}, \bibinfo{author}{{Barkaoui}, K.},
  \bibinfo{author}{{Burgasser}, A.}, \bibinfo{author}{{Burleigh}, M.R.},
  \bibinfo{author}{{Carey}, S.J.}, \bibinfo{author}{{Chaushev}, A.},
  \bibinfo{author}{{Copperwheat}, C.M.}, \bibinfo{author}{{Delrez}, L.},
  \bibinfo{author}{{Fernandes}, C.S.}, \bibinfo{author}{{Holdsworth}, D.L.},
  \bibinfo{author}{{Kotze}, E.J.}, \bibinfo{author}{{Van Grootel}, V.},
  \bibinfo{author}{{Almleaky}, Y.}, \bibinfo{author}{{Benkhaldoun}, Z.},
  \bibinfo{author}{{Magain}, P.}, \bibinfo{author}{{Queloz}, D.},
  \bibinfo{year}{2017}.
\newblock \bibinfo{title}{{Seven temperate terrestrial planets around the
  nearby ultracool dwarf star TRAPPIST-1}}.
\newblock \bibinfo{journal}{\nat} \bibinfo{volume}{542},
  \bibinfo{pages}{456--460}.
\newblock \DOIprefix\doi{10.1038/nature21360},
  \href{http://arxiv.org/abs/1703.01424}{\tt arXiv:1703.01424}.
\bibitem[{{Goldstein}(1950)}]{Goldstein_1950}
\bibinfo{author}{{Goldstein}, H.}, \bibinfo{year}{1950}.
\newblock \bibinfo{title}{{Classical mechanics}}.
\bibitem[{Gomes and Correia(2024)}]{Gomes_Correia_2024p2}
\bibinfo{author}{Gomes, S.R.}, \bibinfo{author}{Correia, A.C.},
  \bibinfo{year}{2024}.
\newblock \bibinfo{title}{Dynamical evolution of the uranian satellite system
  ii. crossing of the 5/3 ariel–umbriel mean motion resonance}.
\newblock \bibinfo{journal}{Icarus} , \bibinfo{pages}{116254}\URLprefix
  \url{https://www.sciencedirect.com/science/article/pii/S0019103524003142},
  \DOIprefix\doi{https://doi.org/10.1016/j.icarus.2024.116254}.
\bibitem[{{Gomes} and {Correia}(2023)}]{Gomes_Correia_2023}
\bibinfo{author}{{Gomes}, S.R.A.}, \bibinfo{author}{{Correia}, A.C.M.},
  \bibinfo{year}{2023}.
\newblock \bibinfo{title}{{Effect of the inclination in the passage through the
  5/3 mean motion resonance between Ariel and Umbriel}}.
\newblock \bibinfo{journal}{\aap} \bibinfo{volume}{674}, \bibinfo{pages}{A111}.
\newblock \DOIprefix\doi{10.1051/0004-6361/202346101},
  \href{http://arxiv.org/abs/2305.08794}{\tt arXiv:2305.08794}.
\bibitem[{{Greenberg}(1975)}]{Greenberg_1975}
\bibinfo{author}{{Greenberg}, R.}, \bibinfo{year}{1975}.
\newblock \bibinfo{title}{{On the Laplace relation among the satellites of
  Uranus.}}
\newblock \bibinfo{journal}{\mnras} \bibinfo{volume}{173},
  \bibinfo{pages}{121--129}.
\newblock \DOIprefix\doi{10.1093/mnras/173.1.121}.
\bibitem[{{Henrard} and {Sato}(1989)}]{Henrard_Sato_1989}
\bibinfo{author}{{Henrard}, J.}, \bibinfo{author}{{Sato}, M.},
  \bibinfo{year}{1989}.
\newblock \bibinfo{title}{{The origin of chaotic behaviour in the
  Miranda-Umbriel 3 : 1 resonances}}.
\newblock \bibinfo{journal}{Celestial Mechanics and Dynamical Astronomy}
  \bibinfo{volume}{47}, \bibinfo{pages}{391--417}.
\newblock \DOIprefix\doi{10.1007/BF00051013}.
\bibitem[{{Hut}(1980)}]{Hut_1980}
\bibinfo{author}{{Hut}, P.}, \bibinfo{year}{1980}.
\newblock \bibinfo{title}{{Stability of tidal equilibrium}}.
\newblock \bibinfo{journal}{\aap} \bibinfo{volume}{92},
  \bibinfo{pages}{167--170}.
\bibitem[{{Ida} et~al.(2020){Ida}, {Ueta}, {Sasaki} and
  {Ishizawa}}]{Ida_etal_2020}
\bibinfo{author}{{Ida}, S.}, \bibinfo{author}{{Ueta}, S.},
  \bibinfo{author}{{Sasaki}, T.}, \bibinfo{author}{{Ishizawa}, Y.},
  \bibinfo{year}{2020}.
\newblock \bibinfo{title}{{Uranian satellite formation by evolution of a water
  vapour disk generated by a giant impact}}.
\newblock \bibinfo{journal}{Nature Astronomy} \bibinfo{volume}{4},
  \bibinfo{pages}{880--885}.
\newblock \DOIprefix\doi{10.1038/s41550-020-1049-8},
  \href{http://arxiv.org/abs/2003.13582}{\tt arXiv:2003.13582}.
\bibitem[{{Inderbitzi} et~al.(2020){Inderbitzi}, {Szul{\'a}gyi}, {Cilibrasi}
  and {Mayer}}]{Inderbitzi_etal_2020}
\bibinfo{author}{{Inderbitzi}, C.}, \bibinfo{author}{{Szul{\'a}gyi}, J.},
  \bibinfo{author}{{Cilibrasi}, M.}, \bibinfo{author}{{Mayer}, L.},
  \bibinfo{year}{2020}.
\newblock \bibinfo{title}{{Formation of satellites in circumplanetary discs
  generated by disc instability}}.
\newblock \bibinfo{journal}{\mnras} \bibinfo{volume}{499},
  \bibinfo{pages}{1023--1036}.
\newblock \DOIprefix\doi{10.1093/mnras/staa2796},
  \href{http://arxiv.org/abs/1912.11406}{\tt arXiv:1912.11406}.
\bibitem[{{Ishizawa} et~al.(2019){Ishizawa}, {Sasaki} and
  {Hosono}}]{Ishizawa_etal_2019}
\bibinfo{author}{{Ishizawa}, Y.}, \bibinfo{author}{{Sasaki}, T.},
  \bibinfo{author}{{Hosono}, N.}, \bibinfo{year}{2019}.
\newblock \bibinfo{title}{{Can the Uranian Satellites Form from a Debris Disk
  Generated by a Giant Impact?}}
\newblock \bibinfo{journal}{\apj} \bibinfo{volume}{885}, \bibinfo{pages}{132}.
\newblock \DOIprefix\doi{10.3847/1538-4357/ab48ef},
  \href{http://arxiv.org/abs/1909.13065}{\tt arXiv:1909.13065}.
\bibitem[{{Jacobson}(2014)}]{Jacobson_2014}
\bibinfo{author}{{Jacobson}, R.A.}, \bibinfo{year}{2014}.
\newblock \bibinfo{title}{{The Orbits of the Uranian Satellites and Rings, the
  Gravity Field of the Uranian System, and the Orientation of the Pole of
  Uranus}}.
\newblock \bibinfo{journal}{\aj} \bibinfo{volume}{148}, \bibinfo{pages}{76}.
\newblock \DOIprefix\doi{10.1088/0004-6256/148/5/76}.
\bibitem[{{Jeffreys}(1976)}]{Jeffreys_1976}
\bibinfo{author}{{Jeffreys}, H.}, \bibinfo{year}{1976}.
\newblock \bibinfo{title}{{The earth. Its origin, history and physical
  constitution.}}
\newblock \bibinfo{publisher}{Cambridge University Press}.
\bibitem[{{Jewitt} and {Haghighipour}(2007)}]{Jewitt_Haghighipour_2007}
\bibinfo{author}{{Jewitt}, D.}, \bibinfo{author}{{Haghighipour}, N.},
  \bibinfo{year}{2007}.
\newblock \bibinfo{title}{{Irregular Satellites of the Planets: Products of
  Capture in the Early Solar System}}.
\newblock \bibinfo{journal}{\araa} \bibinfo{volume}{45},
  \bibinfo{pages}{261--295}.
\newblock \DOIprefix\doi{10.1146/annurev.astro.44.051905.092459},
  \href{http://arxiv.org/abs/astro-ph/0703059}{\tt arXiv:astro-ph/0703059}.
\bibitem[{{Kirchoff} et~al.(2022){Kirchoff}, {Dones}, {Singer} and
  {Schenk}}]{Kirchoff_etal_2022}
\bibinfo{author}{{Kirchoff}, M.R.}, \bibinfo{author}{{Dones}, L.},
  \bibinfo{author}{{Singer}, K.}, \bibinfo{author}{{Schenk}, P.},
  \bibinfo{year}{2022}.
\newblock \bibinfo{title}{{Crater Distributions of Uranus's Mid-sized
  Satellites and Implications for Outer Solar System Bombardment}}.
\newblock \bibinfo{journal}{\psj} \bibinfo{volume}{3}, \bibinfo{pages}{42}.
\newblock \DOIprefix\doi{10.3847/PSJ/ac42d7}.
\bibitem[{{Laskar}(1986)}]{Laskar_1986}
\bibinfo{author}{{Laskar}, J.}, \bibinfo{year}{1986}.
\newblock \bibinfo{title}{{A general theory for the Uranian satellites}}.
\newblock \bibinfo{journal}{\aap} \bibinfo{volume}{166},
  \bibinfo{pages}{349--358}.
\bibitem[{{Laskar}(1990)}]{Laskar_1990}
\bibinfo{author}{{Laskar}, J.}, \bibinfo{year}{1990}.
\newblock \bibinfo{title}{{The chaotic motion of the solar system - A numerical
  estimate of the size of the chaotic zones}}.
\newblock \bibinfo{journal}{\icarus} \bibinfo{volume}{88},
  \bibinfo{pages}{266--291}.
\newblock \DOIprefix\doi{10.1016/0019-1035(90)90084-M}.
\bibitem[{{Laskar}(1993)}]{Laskar_1993}
\bibinfo{author}{{Laskar}, J.}, \bibinfo{year}{1993}.
\newblock \bibinfo{title}{{Frequency analysis for multi-dimensional systems.
  Global dynamics and diffusion}}.
\newblock \bibinfo{journal}{Physica D Nonlinear Phenomena}
  \bibinfo{volume}{67}, \bibinfo{pages}{257--281}.
\newblock \DOIprefix\doi{10.1016/0167-2789(93)90210-R}.
\bibitem[{{Laskar} et~al.(2012){Laskar}, {Bou{\'e}} and
  {Correia}}]{Laskar_etal_2012}
\bibinfo{author}{{Laskar}, J.}, \bibinfo{author}{{Bou{\'e}}, G.},
  \bibinfo{author}{{Correia}, A.C.M.}, \bibinfo{year}{2012}.
\newblock \bibinfo{title}{{Tidal dissipation in multi-planet systems and
  constraints on orbit fitting}}.
\newblock \bibinfo{journal}{\aap} \bibinfo{volume}{538}, \bibinfo{pages}{A105}.
\newblock \DOIprefix\doi{10.1051/0004-6361/201116643},
  \href{http://arxiv.org/abs/1110.4565}{\tt arXiv:1110.4565}.
\bibitem[{{Laskar} and {Jacobson}(1987)}]{Laskar_Jacobson_1987}
\bibinfo{author}{{Laskar}, J.}, \bibinfo{author}{{Jacobson}, R.A.},
  \bibinfo{year}{1987}.
\newblock \bibinfo{title}{{GUST86 - an analytical ephemeris of the Uranian
  satellites}}.
\newblock \bibinfo{journal}{\aap} \bibinfo{volume}{188},
  \bibinfo{pages}{212--224}.
\bibitem[{{Laskar} and {Robutel}(1993)}]{Laskar_Robutel_1993}
\bibinfo{author}{{Laskar}, J.}, \bibinfo{author}{{Robutel}, P.},
  \bibinfo{year}{1993}.
\newblock \bibinfo{title}{{The chaotic obliquity of the planets}}.
\newblock \bibinfo{journal}{Nature} \bibinfo{volume}{361},
  \bibinfo{pages}{608--612}.
\newblock \DOIprefix\doi{10.1038/361608a0}.
\bibitem[{{Leleu} et~al.(2021){Leleu}, {Alibert}, {Hara}, {Hooton}, {Wilson},
  {Robutel}, {Delisle}, {Laskar}, {Hoyer}, {Lovis}, {Bryant}, {Ducrot},
  {Cabrera}, {Delrez}, {Acton}, {Adibekyan}, {Allart}, {Allende Prieto},
  {Alonso}, {Alves}, {Anderson}, {Angerhausen}, {Anglada Escud{\'e}},
  {Asquier}, {Barrado}, {Barros}, {Baumjohann}, {Bayliss}, {Beck}, {Beck},
  {Bekkelien}, {Benz}, {Billot}, {Bonfanti}, {Bonfils}, {Bouchy}, {Bourrier},
  {Bou{\'e}}, {Brandeker}, {Broeg}, {Buder}, {Burdanov}, {Burleigh},
  {B{\'a}rczy}, {Cameron}, {Chamberlain}, {Charnoz}, {Cooke}, {Corral Van
  Damme}, {Correia}, {Cristiani}, {Damasso}, {Davies}, {Deleuil}, {Demangeon},
  {Demory}, {Di Marcantonio}, {Di Persio}, {Dumusque}, {Ehrenreich}, {Erikson},
  {Figueira}, {Fortier}, {Fossati}, {Fridlund}, {Futyan}, {Gandolfi},
  {Garc{\'\i}a Mu{\~n}oz}, {Garcia}, {Gill}, {Gillen}, {Gillon}, {Goad},
  {Gonz{\'a}lez Hern{\'a}ndez}, {Guedel}, {G{\"u}nther}, {Haldemann},
  {Henderson}, {Heng}, {Hogan}, {Isaak}, {Jehin}, {Jenkins}, {Jord{\'a}n},
  {Kiss}, {Kristiansen}, {Lam}, {Lavie}, {Lecavelier des Etangs}, {Lendl},
  {Lillo-Box}, {Lo Curto}, {Magrin}, {Martins}, {Maxted}, {McCormac}, {Mehner},
  {Micela}, {Molaro}, {Moyano}, {Murray}, {Nascimbeni}, {Nunes}, {Olofsson},
  {Osborn}, {Oshagh}, {Ottensamer}, {Pagano}, {Pall{\'e}}, {Pedersen}, {Pepe},
  {Persson}, {Peter}, {Piotto}, {Polenta}, {Pollacco}, {Poretti}, {Pozuelos},
  {Queloz}, {Ragazzoni}, {Rando}, {Ratti}, {Rauer}, {Raynard}, {Rebolo},
  {Reimers}, {Ribas}, {Santos}, {Scandariato}, {Schneider}, {Sebastian},
  {Sestovic}, {Simon}, {Smith}, {Sousa}, {Sozzetti}, {Steller}, {Su{\'a}rez
  Mascare{\~n}o}, {Szab{\'o}}, {S{\'e}gransan}, {Thomas}, {Thompson},
  {Tilbrook}, {Triaud}, {Turner}, {Udry}, {Van Grootel}, {Venus}, {Verrecchia},
  {Vines}, {Walton}, {West}, {Wheatley}, {Wolter} and {Zapatero
  Osorio}}]{Leleu_etal_2021}
\bibinfo{author}{{Leleu}, A.}, \bibinfo{author}{{Alibert}, Y.},
  \bibinfo{author}{{Hara}, N.C.}, \bibinfo{author}{{Hooton}, M.J.},
  \bibinfo{author}{{Wilson}, T.G.}, \bibinfo{author}{{Robutel}, P.},
  \bibinfo{author}{{Delisle}, J.B.}, \bibinfo{author}{{Laskar}, J.},
  \bibinfo{author}{{Hoyer}, S.}, \bibinfo{author}{{Lovis}, C.},
  \bibinfo{author}{{Bryant}, E.M.}, \bibinfo{author}{{Ducrot}, E.},
  \bibinfo{author}{{Cabrera}, J.}, \bibinfo{author}{{Delrez}, L.},
  \bibinfo{author}{{Acton}, J.S.}, \bibinfo{author}{{Adibekyan}, V.},
  \bibinfo{author}{{Allart}, R.}, \bibinfo{author}{{Allende Prieto}, C.},
  \bibinfo{author}{{Alonso}, R.}, \bibinfo{author}{{Alves}, D.},
  \bibinfo{author}{{Anderson}, D.R.}, \bibinfo{author}{{Angerhausen}, D.},
  \bibinfo{author}{{Anglada Escud{\'e}}, G.}, \bibinfo{author}{{Asquier}, J.},
  \bibinfo{author}{{Barrado}, D.}, \bibinfo{author}{{Barros}, S.C.C.},
  \bibinfo{author}{{Baumjohann}, W.}, \bibinfo{author}{{Bayliss}, D.},
  \bibinfo{author}{{Beck}, M.}, \bibinfo{author}{{Beck}, T.},
  \bibinfo{author}{{Bekkelien}, A.}, \bibinfo{author}{{Benz}, W.},
  \bibinfo{author}{{Billot}, N.}, \bibinfo{author}{{Bonfanti}, A.},
  \bibinfo{author}{{Bonfils}, X.}, \bibinfo{author}{{Bouchy}, F.},
  \bibinfo{author}{{Bourrier}, V.}, \bibinfo{author}{{Bou{\'e}}, G.},
  \bibinfo{author}{{Brandeker}, A.}, \bibinfo{author}{{Broeg}, C.},
  \bibinfo{author}{{Buder}, M.}, \bibinfo{author}{{Burdanov}, A.},
  \bibinfo{author}{{Burleigh}, M.R.}, \bibinfo{author}{{B{\'a}rczy}, T.},
  \bibinfo{author}{{Cameron}, A.C.}, \bibinfo{author}{{Chamberlain}, S.},
  \bibinfo{author}{{Charnoz}, S.}, \bibinfo{author}{{Cooke}, B.F.},
  \bibinfo{author}{{Corral Van Damme}, C.}, \bibinfo{author}{{Correia},
  A.C.M.}, \bibinfo{author}{{Cristiani}, S.}, \bibinfo{author}{{Damasso}, M.},
  \bibinfo{author}{{Davies}, M.B.}, \bibinfo{author}{{Deleuil}, M.},
  \bibinfo{author}{{Demangeon}, O.D.S.}, \bibinfo{author}{{Demory}, B.O.},
  \bibinfo{author}{{Di Marcantonio}, P.}, \bibinfo{author}{{Di Persio}, G.},
  \bibinfo{author}{{Dumusque}, X.}, \bibinfo{author}{{Ehrenreich}, D.},
  \bibinfo{author}{{Erikson}, A.}, \bibinfo{author}{{Figueira}, P.},
  \bibinfo{author}{{Fortier}, A.}, \bibinfo{author}{{Fossati}, L.},
  \bibinfo{author}{{Fridlund}, M.}, \bibinfo{author}{{Futyan}, D.},
  \bibinfo{author}{{Gandolfi}, D.}, \bibinfo{author}{{Garc{\'\i}a Mu{\~n}oz},
  A.}, \bibinfo{author}{{Garcia}, L.J.}, \bibinfo{author}{{Gill}, S.},
  \bibinfo{author}{{Gillen}, E.}, \bibinfo{author}{{Gillon}, M.},
  \bibinfo{author}{{Goad}, M.R.}, \bibinfo{author}{{Gonz{\'a}lez
  Hern{\'a}ndez}, J.I.}, \bibinfo{author}{{Guedel}, M.},
  \bibinfo{author}{{G{\"u}nther}, M.N.}, \bibinfo{author}{{Haldemann}, J.},
  \bibinfo{author}{{Henderson}, B.}, \bibinfo{author}{{Heng}, K.},
  \bibinfo{author}{{Hogan}, A.E.}, \bibinfo{author}{{Isaak}, K.},
  \bibinfo{author}{{Jehin}, E.}, \bibinfo{author}{{Jenkins}, J.S.},
  \bibinfo{author}{{Jord{\'a}n}, A.}, \bibinfo{author}{{Kiss}, L.},
  \bibinfo{author}{{Kristiansen}, M.H.}, \bibinfo{author}{{Lam}, K.},
  \bibinfo{author}{{Lavie}, B.}, \bibinfo{author}{{Lecavelier des Etangs}, A.},
  \bibinfo{author}{{Lendl}, M.}, \bibinfo{author}{{Lillo-Box}, J.},
  \bibinfo{author}{{Lo Curto}, G.}, \bibinfo{author}{{Magrin}, D.},
  \bibinfo{author}{{Martins}, C.J.A.P.}, \bibinfo{author}{{Maxted}, P.F.L.},
  \bibinfo{author}{{McCormac}, J.}, \bibinfo{author}{{Mehner}, A.},
  \bibinfo{author}{{Micela}, G.}, \bibinfo{author}{{Molaro}, P.},
  \bibinfo{author}{{Moyano}, M.}, \bibinfo{author}{{Murray}, C.A.},
  \bibinfo{author}{{Nascimbeni}, V.}, \bibinfo{author}{{Nunes}, N.J.},
  \bibinfo{author}{{Olofsson}, G.}, \bibinfo{author}{{Osborn}, H.P.},
  \bibinfo{author}{{Oshagh}, M.}, \bibinfo{author}{{Ottensamer}, R.},
  \bibinfo{author}{{Pagano}, I.}, \bibinfo{author}{{Pall{\'e}}, E.},
  \bibinfo{author}{{Pedersen}, P.P.}, \bibinfo{author}{{Pepe}, F.A.},
  \bibinfo{author}{{Persson}, C.M.}, \bibinfo{author}{{Peter}, G.},
  \bibinfo{author}{{Piotto}, G.}, \bibinfo{author}{{Polenta}, G.},
  \bibinfo{author}{{Pollacco}, D.}, \bibinfo{author}{{Poretti}, E.},
  \bibinfo{author}{{Pozuelos}, F.J.}, \bibinfo{author}{{Queloz}, D.},
  \bibinfo{author}{{Ragazzoni}, R.}, \bibinfo{author}{{Rando}, N.},
  \bibinfo{author}{{Ratti}, F.}, \bibinfo{author}{{Rauer}, H.},
  \bibinfo{author}{{Raynard}, L.}, \bibinfo{author}{{Rebolo}, R.},
  \bibinfo{author}{{Reimers}, C.}, \bibinfo{author}{{Ribas}, I.},
  \bibinfo{author}{{Santos}, N.C.}, \bibinfo{author}{{Scandariato}, G.},
  \bibinfo{author}{{Schneider}, J.}, \bibinfo{author}{{Sebastian}, D.},
  \bibinfo{author}{{Sestovic}, M.}, \bibinfo{author}{{Simon}, A.E.},
  \bibinfo{author}{{Smith}, A.M.S.}, \bibinfo{author}{{Sousa}, S.G.},
  \bibinfo{author}{{Sozzetti}, A.}, \bibinfo{author}{{Steller}, M.},
  \bibinfo{author}{{Su{\'a}rez Mascare{\~n}o}, A.},
  \bibinfo{author}{{Szab{\'o}}, G.M.}, \bibinfo{author}{{S{\'e}gransan}, D.},
  \bibinfo{author}{{Thomas}, N.}, \bibinfo{author}{{Thompson}, S.},
  \bibinfo{author}{{Tilbrook}, R.H.}, \bibinfo{author}{{Triaud}, A.},
  \bibinfo{author}{{Turner}, O.}, \bibinfo{author}{{Udry}, S.},
  \bibinfo{author}{{Van Grootel}, V.}, \bibinfo{author}{{Venus}, H.},
  \bibinfo{author}{{Verrecchia}, F.}, \bibinfo{author}{{Vines}, J.I.},
  \bibinfo{author}{{Walton}, N.A.}, \bibinfo{author}{{West}, R.G.},
  \bibinfo{author}{{Wheatley}, P.J.}, \bibinfo{author}{{Wolter}, D.},
  \bibinfo{author}{{Zapatero Osorio}, M.R.}, \bibinfo{year}{2021}.
\newblock \bibinfo{title}{{Six transiting planets and a chain of Laplace
  resonances in TOI-178}}.
\newblock \bibinfo{journal}{\aap} \bibinfo{volume}{649}, \bibinfo{pages}{A26}.
\newblock \DOIprefix\doi{10.1051/0004-6361/202039767},
  \href{http://arxiv.org/abs/2101.09260}{\tt arXiv:2101.09260}.
\bibitem[{{Levrard} et~al.(2007){Levrard}, {Correia}, {Chabrier}, {Baraffe},
  {Selsis} and {Laskar}}]{Levrard_etal_2007}
\bibinfo{author}{{Levrard}, B.}, \bibinfo{author}{{Correia}, A.C.M.},
  \bibinfo{author}{{Chabrier}, G.}, \bibinfo{author}{{Baraffe}, I.},
  \bibinfo{author}{{Selsis}, F.}, \bibinfo{author}{{Laskar}, J.},
  \bibinfo{year}{2007}.
\newblock \bibinfo{title}{{Tidal dissipation within hot Jupiters: a new
  appraisal}}.
\newblock \bibinfo{journal}{\aap} \bibinfo{volume}{462},
  \bibinfo{pages}{L5--L8}.
\newblock \DOIprefix\doi{10.1051/0004-6361:20066487},
  \href{http://arxiv.org/abs/astro-ph/0612044}{\tt arXiv:astro-ph/0612044}.
\bibitem[{{Lissauer} et~al.(2011){Lissauer}, {Ragozzine}, {Fabrycky},
  {Steffen}, {Ford}, {Jenkins}, {Shporer}, {Holman}, {Rowe}, {Quintana},
  {Batalha}, {Borucki}, {Bryson}, {Caldwell}, {Carter}, {Ciardi}, {Dunham},
  {Fortney}, {Gautier}, {Howell}, {Koch}, {Latham}, {Marcy}, {Morehead} and
  {Sasselov}}]{Lissauer_etal_2011}
\bibinfo{author}{{Lissauer}, J.J.}, \bibinfo{author}{{Ragozzine}, D.},
  \bibinfo{author}{{Fabrycky}, D.C.}, \bibinfo{author}{{Steffen}, J.H.},
  \bibinfo{author}{{Ford}, E.B.}, \bibinfo{author}{{Jenkins}, J.M.},
  \bibinfo{author}{{Shporer}, A.}, \bibinfo{author}{{Holman}, M.J.},
  \bibinfo{author}{{Rowe}, J.F.}, \bibinfo{author}{{Quintana}, E.V.},
  \bibinfo{author}{{Batalha}, N.M.}, \bibinfo{author}{{Borucki}, W.J.},
  \bibinfo{author}{{Bryson}, S.T.}, \bibinfo{author}{{Caldwell}, D.A.},
  \bibinfo{author}{{Carter}, J.A.}, \bibinfo{author}{{Ciardi}, D.},
  \bibinfo{author}{{Dunham}, E.W.}, \bibinfo{author}{{Fortney}, J.J.},
  \bibinfo{author}{{Gautier}, Thomas~N., I.}, \bibinfo{author}{{Howell}, S.B.},
  \bibinfo{author}{{Koch}, D.G.}, \bibinfo{author}{{Latham}, D.W.},
  \bibinfo{author}{{Marcy}, G.W.}, \bibinfo{author}{{Morehead}, R.C.},
  \bibinfo{author}{{Sasselov}, D.}, \bibinfo{year}{2011}.
\newblock \bibinfo{title}{{Architecture and Dynamics of Kepler's Candidate
  Multiple Transiting Planet Systems}}.
\newblock \bibinfo{journal}{\apjs} \bibinfo{volume}{197}, \bibinfo{pages}{8}.
\newblock \DOIprefix\doi{10.1088/0067-0049/197/1/8},
  \href{http://arxiv.org/abs/1102.0543}{\tt arXiv:1102.0543}.
\bibitem[{{Malhotra} and {Dermott}(1990)}]{Malhotra_Dermott_1990}
\bibinfo{author}{{Malhotra}, R.}, \bibinfo{author}{{Dermott}, S.F.},
  \bibinfo{year}{1990}.
\newblock \bibinfo{title}{{The role of secondary resonances in the orbital
  history of Miranda}}.
\newblock \bibinfo{journal}{\icarus} \bibinfo{volume}{85},
  \bibinfo{pages}{444--480}.
\newblock \DOIprefix\doi{10.1016/0019-1035(90)90126-T}.
\bibitem[{{Malhotra} et~al.(1989){Malhotra}, {Fox}, {Murray} and
  {Nicholson}}]{Malhotra_etal_1989}
\bibinfo{author}{{Malhotra}, R.}, \bibinfo{author}{{Fox}, K.},
  \bibinfo{author}{{Murray}, C.D.}, \bibinfo{author}{{Nicholson}, P.D.},
  \bibinfo{year}{1989}.
\newblock \bibinfo{title}{{Secular perturbations of the Uranian satellites:
  theory and practice.}}
\newblock \bibinfo{journal}{\aap} \bibinfo{volume}{221},
  \bibinfo{pages}{348--358}.
\bibitem[{{Mardling}(2007)}]{Mardling_2007}
\bibinfo{author}{{Mardling}, R.A.}, \bibinfo{year}{2007}.
\newblock \bibinfo{title}{{Long-term tidal evolution of short-period planets
  with companions}}.
\newblock \bibinfo{journal}{\mnras} \bibinfo{volume}{382},
  \bibinfo{pages}{1768--1790}.
\newblock \DOIprefix\doi{10.1111/j.1365-2966.2007.12500.x},
  \href{http://arxiv.org/abs/0706.0224}{\tt arXiv:0706.0224}.
\bibitem[{{Mignard}(1979)}]{Mignard_1979}
\bibinfo{author}{{Mignard}, F.}, \bibinfo{year}{1979}.
\newblock \bibinfo{title}{{The Evolution of the Lunar Orbit Revisited. I.}}
\newblock \bibinfo{journal}{Moon and Planets} \bibinfo{volume}{20},
  \bibinfo{pages}{301--315}.
\newblock \DOIprefix\doi{10.1007/BF00907581}.
\bibitem[{{Milani} et~al.(1987){Milani}, {Nobili} and
  {Carpino}}]{Milani_etal_1987}
\bibinfo{author}{{Milani}, A.}, \bibinfo{author}{{Nobili}, A.M.},
  \bibinfo{author}{{Carpino}, M.}, \bibinfo{year}{1987}.
\newblock \bibinfo{title}{{Secular variations of the semimajor axes - Theory
  and experiments}}.
\newblock \bibinfo{journal}{\aap} \bibinfo{volume}{172},
  \bibinfo{pages}{265--279}.
\bibitem[{{Murray} and {Dermott}(1999)}]{Murray_Dermott_1999}
\bibinfo{author}{{Murray}, C.D.}, \bibinfo{author}{{Dermott}, S.F.},
  \bibinfo{year}{1999}.
\newblock \bibinfo{title}{{Solar system dynamics}}.
\newblock \bibinfo{publisher}{Cambridge University Press}.
\bibitem[{{Ogilvie} and {Lin}(2007)}]{Ogilvie_Lin_2007}
\bibinfo{author}{{Ogilvie}, G.I.}, \bibinfo{author}{{Lin}, D.N.C.},
  \bibinfo{year}{2007}.
\newblock \bibinfo{title}{{Tidal Dissipation in Rotating Solar-Type Stars}}.
\newblock \bibinfo{journal}{\apj} \bibinfo{volume}{661},
  \bibinfo{pages}{1180--1191}.
\newblock \DOIprefix\doi{10.1086/515435},
  \href{http://arxiv.org/abs/astro-ph/0702492}{\tt arXiv:astro-ph/0702492}.
\bibitem[{{Peale}(1988)}]{Peale_1988}
\bibinfo{author}{{Peale}, S.J.}, \bibinfo{year}{1988}.
\newblock \bibinfo{title}{{Speculative histories of the Uranian satellite
  system}}.
\newblock \bibinfo{journal}{\icarus} \bibinfo{volume}{74},
  \bibinfo{pages}{153--171}.
\newblock \DOIprefix\doi{10.1016/0019-1035(88)90037-1}.
\bibitem[{{Peale}(1999)}]{Peale_1999}
\bibinfo{author}{{Peale}, S.J.}, \bibinfo{year}{1999}.
\newblock \bibinfo{title}{{Origin and Evolution of the Natural Satellites}}.
\newblock \bibinfo{journal}{\araa} \bibinfo{volume}{37},
  \bibinfo{pages}{533--602}.
\newblock \DOIprefix\doi{10.1146/annurev.astro.37.1.533}.
\bibitem[{{Plescia}(1987)}]{Plescia_1987}
\bibinfo{author}{{Plescia}, J.B.}, \bibinfo{year}{1987}.
\newblock \bibinfo{title}{{Cratering history of the Uranian satellites:
  Umbriel, Titania, and Oberon}}.
\newblock \bibinfo{journal}{\jgr} \bibinfo{volume}{92},
  \bibinfo{pages}{14918--14932}.
\newblock \DOIprefix\doi{10.1029/JA092iA13p14918}.
\bibitem[{{Pollack} et~al.(1991){Pollack}, {Lunine} and
  {Tittemore}}]{Pollack_etal_1991}
\bibinfo{author}{{Pollack}, J.B.}, \bibinfo{author}{{Lunine}, J.I.},
  \bibinfo{author}{{Tittemore}, W.C.}, \bibinfo{year}{1991}.
\newblock \bibinfo{title}{{Origin of the Uranian satellites.}}, in:
  \bibinfo{editor}{{Bergstralh}, J.T.}, \bibinfo{editor}{{Miner}, E.D.},
  \bibinfo{editor}{{Matthews}, M.S.} (Eds.), \bibinfo{booktitle}{Uranus}, pp.
  \bibinfo{pages}{469--512}.
\bibitem[{{Prentice}(1986)}]{Prentice_1986}
\bibinfo{author}{{Prentice}, A.J.R.}, \bibinfo{year}{1986}.
\newblock \bibinfo{title}{{Uranus after Voyager 2 and the origin of the solar
  system}}.
\newblock \bibinfo{journal}{Proceedings of the Astronomical Society of
  Australia} \bibinfo{volume}{6}, \bibinfo{pages}{394--402}.
\bibitem[{{Rogoszinski} and {Hamilton}(2021)}]{Rogoszinski_Hamilton_2020}
\bibinfo{author}{{Rogoszinski}, Z.}, \bibinfo{author}{{Hamilton}, D.P.},
  \bibinfo{year}{2021}.
\newblock \bibinfo{title}{{Tilting Uranus: Collisions versus Spin-Orbit
  Resonance}}.
\newblock \bibinfo{journal}{\psj} \bibinfo{volume}{2}, \bibinfo{pages}{78}.
\newblock \DOIprefix\doi{10.3847/PSJ/abec4e},
  \href{http://arxiv.org/abs/2004.14913}{\tt arXiv:2004.14913}.
\bibitem[{{Rufu} and {Canup}(2022)}]{Rufu_Canup_2022}
\bibinfo{author}{{Rufu}, R.}, \bibinfo{author}{{Canup}, R.M.},
  \bibinfo{year}{2022}.
\newblock \bibinfo{title}{{Coaccretion + Giant-impact Origin of the Uranus
  System: Tilting Impact}}.
\newblock \bibinfo{journal}{\apj} \bibinfo{volume}{928}, \bibinfo{pages}{123}.
\newblock \DOIprefix\doi{10.3847/1538-4357/ac525a},
  \href{http://arxiv.org/abs/2204.00124}{\tt arXiv:2204.00124}.
\bibitem[{{Singer}(1968)}]{Singer_1968}
\bibinfo{author}{{Singer}, S.F.}, \bibinfo{year}{1968}.
\newblock \bibinfo{title}{{The Origin of the Moon and Geophysical
  Consequences}}.
\newblock \bibinfo{journal}{Geophysical Journal International}
  \bibinfo{volume}{15}, \bibinfo{pages}{191--204}.
\newblock \DOIprefix\doi{10.1111/j.1365-246X.1968.tb05758.x}.
\bibitem[{{Smart}(1965)}]{Smart_1965}
\bibinfo{author}{{Smart}, W.M.}, \bibinfo{year}{1965}.
\newblock \bibinfo{title}{{Text-book on spherical astronomy}}.
\newblock \bibinfo{publisher}{Cambridge University Press}.
\bibitem[{{Smith} et~al.(1986){Smith}, {Soderblom}, {Beebe}, {Bliss}, {Boyce},
  {Brahic}, {Briggs}, {Brown}, {Collins}, {Cook}, {Croft}, {Cuzzi},
  {Danielson}, {Davies}, {Dowling}, {Godfrey}, {Hansen}, {Harris}, {Hunt},
  {Ingersoll}, {Johnson}, {Krauss}, {Masursky}, {Morrison}, {Owen}, {Plescia},
  {Pollack}, {Porco}, {Rages}, {Sagan}, {Shoemaker}, {Sromovsky}, {Stoker},
  {Strom}, {Suomi}, {Synnott}, {Terrile}, {Thomas}, {Thompson} and
  {Veverka}}]{Smith_etal_1986}
\bibinfo{author}{{Smith}, B.A.}, \bibinfo{author}{{Soderblom}, L.A.},
  \bibinfo{author}{{Beebe}, R.}, \bibinfo{author}{{Bliss}, D.},
  \bibinfo{author}{{Boyce}, J.M.}, \bibinfo{author}{{Brahic}, A.},
  \bibinfo{author}{{Briggs}, G.A.}, \bibinfo{author}{{Brown}, R.H.},
  \bibinfo{author}{{Collins}, S.A.}, \bibinfo{author}{{Cook}, A.F.},
  \bibinfo{author}{{Croft}, S.K.}, \bibinfo{author}{{Cuzzi}, J.N.},
  \bibinfo{author}{{Danielson}, G.E.}, \bibinfo{author}{{Davies}, M.E.},
  \bibinfo{author}{{Dowling}, T.E.}, \bibinfo{author}{{Godfrey}, D.},
  \bibinfo{author}{{Hansen}, C.J.}, \bibinfo{author}{{Harris}, C.},
  \bibinfo{author}{{Hunt}, G.E.}, \bibinfo{author}{{Ingersoll}, A.P.},
  \bibinfo{author}{{Johnson}, T.V.}, \bibinfo{author}{{Krauss}, R.J.},
  \bibinfo{author}{{Masursky}, H.}, \bibinfo{author}{{Morrison}, D.},
  \bibinfo{author}{{Owen}, T.}, \bibinfo{author}{{Plescia}, J.B.},
  \bibinfo{author}{{Pollack}, J.B.}, \bibinfo{author}{{Porco}, C.C.},
  \bibinfo{author}{{Rages}, K.}, \bibinfo{author}{{Sagan}, C.},
  \bibinfo{author}{{Shoemaker}, E.M.}, \bibinfo{author}{{Sromovsky}, L.A.},
  \bibinfo{author}{{Stoker}, C.}, \bibinfo{author}{{Strom}, R.G.},
  \bibinfo{author}{{Suomi}, V.E.}, \bibinfo{author}{{Synnott}, S.P.},
  \bibinfo{author}{{Terrile}, R.J.}, \bibinfo{author}{{Thomas}, P.},
  \bibinfo{author}{{Thompson}, W.R.}, \bibinfo{author}{{Veverka}, J.},
  \bibinfo{year}{1986}.
\newblock \bibinfo{title}{{Voyager 2 in the Uranian System: Imaging Science
  Results}}.
\newblock \bibinfo{journal}{Science} \bibinfo{volume}{233},
  \bibinfo{pages}{43--64}.
\newblock \DOIprefix\doi{10.1126/science.233.4759.43}.
\bibitem[{{Squyres} et~al.(1985){Squyres}, {Reynolds} and
  {Lissauer}}]{Squyres_etal_1985}
\bibinfo{author}{{Squyres}, S.W.}, \bibinfo{author}{{Reynolds}, R.T.},
  \bibinfo{author}{{Lissauer}, J.J.}, \bibinfo{year}{1985}.
\newblock \bibinfo{title}{{The enigma of the Uranian satellites' orbital
  eccentricities}}.
\newblock \bibinfo{journal}{\icarus} \bibinfo{volume}{61},
  \bibinfo{pages}{218--223}.
\newblock \DOIprefix\doi{10.1016/0019-1035(85)90103-4}.
\bibitem[{{Szul{\'a}gyi} et~al.(2018){Szul{\'a}gyi}, {Cilibrasi} and
  {Mayer}}]{Szulagyi_etal_2018}
\bibinfo{author}{{Szul{\'a}gyi}, J.}, \bibinfo{author}{{Cilibrasi}, M.},
  \bibinfo{author}{{Mayer}, L.}, \bibinfo{year}{2018}.
\newblock \bibinfo{title}{{In Situ Formation of Icy Moons of Uranus and
  Neptune}}.
\newblock \bibinfo{journal}{\apjl} \bibinfo{volume}{868}, \bibinfo{pages}{L13}.
\newblock \DOIprefix\doi{10.3847/2041-8213/aaeed6},
  \href{http://arxiv.org/abs/1811.06574}{\tt arXiv:1811.06574}.
\bibitem[{{Thomas}(1988)}]{Thomas_1988}
\bibinfo{author}{{Thomas}, P.C.}, \bibinfo{year}{1988}.
\newblock \bibinfo{title}{{Radii, shapes, and topography of the satellites of
  Uranus from limb coordinates}}.
\newblock \bibinfo{journal}{\icarus} \bibinfo{volume}{73},
  \bibinfo{pages}{427--441}.
\newblock \DOIprefix\doi{10.1016/0019-1035(88)90054-1}.
\bibitem[{{Tittemore} and {Wisdom}(1988)}]{Tittemore_Wisdom_1988}
\bibinfo{author}{{Tittemore}, W.C.}, \bibinfo{author}{{Wisdom}, J.},
  \bibinfo{year}{1988}.
\newblock \bibinfo{title}{{Tidal evolution of the Uranian satellites I. Passage
  of Ariel and Umbriel through the 5:3 mean-motion commensurability}}.
\newblock \bibinfo{journal}{\icarus} \bibinfo{volume}{74},
  \bibinfo{pages}{172--230}.
\newblock \DOIprefix\doi{10.1016/0019-1035(88)90038-3}.
\bibitem[{{Tittemore} and {Wisdom}(1989)}]{Tittemore_Wisdom_1989}
\bibinfo{author}{{Tittemore}, W.C.}, \bibinfo{author}{{Wisdom}, J.},
  \bibinfo{year}{1989}.
\newblock \bibinfo{title}{{Tidal evolution of the Uranian satellites II. An
  explanation of the anomalously high orbital inclination of Miranda}}.
\newblock \bibinfo{journal}{\icarus} \bibinfo{volume}{78},
  \bibinfo{pages}{63--89}.
\newblock \DOIprefix\doi{10.1016/0019-1035(89)90070-5}.
\bibitem[{{Tittemore} and {Wisdom}(1990)}]{Tittemore_Wisdom_1990}
\bibinfo{author}{{Tittemore}, W.C.}, \bibinfo{author}{{Wisdom}, J.},
  \bibinfo{year}{1990}.
\newblock \bibinfo{title}{{Tidal evolution of the Uranian satellites III.
  Evolution through the Miranda-Umbriel 3:1, Miranda-Ariel 5:3, and
  Ariel-Umbriel 2:1 mean-motion commensurabilities}}.
\newblock \bibinfo{journal}{\icarus} \bibinfo{volume}{85},
  \bibinfo{pages}{394--443}.
\newblock \DOIprefix\doi{10.1016/0019-1035(90)90125-S}.
\bibitem[{{Verheylewegen} et~al.(2013){Verheylewegen}, {Noyelles} and
  {Lemaitre}}]{Verheylewegen_etal_2013}
\bibinfo{author}{{Verheylewegen}, E.}, \bibinfo{author}{{Noyelles}, B.},
  \bibinfo{author}{{Lemaitre}, A.}, \bibinfo{year}{2013}.
\newblock \bibinfo{title}{{A numerical exploration of Miranda's dynamical
  history}}.
\newblock \bibinfo{journal}{\mnras} \bibinfo{volume}{435},
  \bibinfo{pages}{1776--1787}.
\newblock \DOIprefix\doi{10.1093/mnras/stt1415},
  \href{http://arxiv.org/abs/1302.4329}{\tt arXiv:1302.4329}.
\bibitem[{{Ward}(1975)}]{Ward_1975}
\bibinfo{author}{{Ward}, W.R.}, \bibinfo{year}{1975}.
\newblock \bibinfo{title}{{Tidal friction and generalized Cassini's laws in the
  solar system}}.
\newblock \bibinfo{journal}{\aj} \bibinfo{volume}{80}, \bibinfo{pages}{64--70}.
\bibitem[{{Ward} and {Hamilton}(2004)}]{Ward_Hamilton_2004}
\bibinfo{author}{{Ward}, W.R.}, \bibinfo{author}{{Hamilton}, D.P.},
  \bibinfo{year}{2004}.
\newblock \bibinfo{title}{{Tilting Saturn. I. Analytic Model}}.
\newblock \bibinfo{journal}{\aj} \bibinfo{volume}{128},
  \bibinfo{pages}{2501--2509}.
\newblock \DOIprefix\doi{10.1086/424533}.

\end{thebibliography}

\end{document}